\documentclass[twocolumn]{aastex62}

\defcitealias{K15}{K15}
\defcitealias{Guo2013PhotoCat}{G13}

\begin{document}

\title{\MakeUppercase{CANDELS Meets GSWLC: Evolution of the Relationship Between Morphology and Star Formation Since \lowercase{z} = 2}} %\footnote{Released on January, 8th, 2018}

\author{Chandler Osborne}
\affil{Department of Astronomy, Indiana University, Bloomington, IN, 47408}

\author{Samir Salim}
\affil{Department of Astronomy, Indiana University, Bloomington, IN, 47408}

\author{Ivana Damjanov}
\affil{Department of Astronomy and Physics, Saint Mary's University, 923 Robie Street, Halifax, NS B3H 3C3, Canada}

\author{S.M. Faber}
\affil{Department of Astronomy and Astrophysics, University of California, Santa Cruz, CA, 95064}

\author{Marc Huertas-Company}
\affil{GEPI, Observatoire de Paris, CNRS, Universit\'e Paris, 75014 Paris, France}

\author{David C. Koo}
\affil{Department of Astronomy and Astrophysics, University of California, Santa Cruz, CA, 95064}

\author{Kameswara Bharadwaj Mantha}
\affil{Department of Physics and Astronomy, University of Missouri-Kansas City, Kansas City, MO 64110, USA}

\author{Daniel H. McIntosh}
\affil{Department of Physics and Astronomy, University of Missouri-Kansas City, Kansas City, MO 64110, USA}

\author{Joel R. Primack}
\affil{Physics Department, University of California, Santa Cruz, CA 95064, USA}

\author{Sandro Tacchella}
\affil{Center for Astrophysics $\vert$ Harvard $\&$ Smithsonian, 60 Garden St, Cambridge, MA 02138, USA}

\begin{abstract}

Galaxy morphology and its evolution over the cosmic epoch hold important clues for understanding the regulation of star formation (SF).  However, studying the relationship between morphology and SF has been hindered by the availability of consistent data at different redshifts.  Our sample, combining CANDELS ($0.8 < z < 2.5$) and the \textit{GALEX}-SDSS-WISE Legacy Catalog (GSWLC; $z \sim 0$), has physical parameters derived using consistent SED fitting with flexible dust attenuation laws.  We adopt visual classifications from \citet{K15} and expand them to $z\sim0$ using SDSS images matching the physical resolution of CANDELS rest-frame optical images and deep FUV \textit{GALEX} images matching the physical resolution of the CANDELS rest-frame FUV images.  Our main finding is that disks with SF clumps at $z \sim 0$ make a similar fraction ($\sim 15\%$) of star-forming galaxies as at $z\sim2$.  The clumpy disk contribution to the SF budget peaks at $z \sim 1$, rather than $z \sim 2$, suggesting that the principal epoch of disk assembly continues to lower redshifts.  Star-forming spheroids (``blue nuggets''), though less centrally concentrated than quenched spheroids, contribute significantly ($\sim 15\%$) to the SF budget at $z \sim1$--2, suggesting that compaction precedes quenching.  Among green valley and quiescent galaxies, the pure spheroid fraction drops since $z\sim1$, whereas spheroids with disks (S0-like) become dominant.  Mergers at or nearing coalescence are enhanced in SFR relative to the main sequence at all redshifts by a factor of $\sim2$, but contribute $\lesssim 5\%$ to the SF budget, with their contribution remaining small above the main sequence.  

% Among low-sSFR galaxies off the MS, 

\end{abstract}

\section{Introduction}

The classification of galaxies based on their visual appearance (morphology) has its roots in the `tuning fork' diagram \citep{Hubble1926}.  The morphology of a galaxy is known to be correlated with intrinsic properties such as stellar mass, specific star formation rate (sSFR) and color, as well as external properties such as environment \citep{Dressler1980MorphDensityRelation, Roberts&Haynes1994PhysicalParams&Morphology, Kennicutt1998HubbleSequence&SFR, GildePaz2007GALEXUVAtlas, Pozzetti2010GSMFEvolution, Bait2017Morphology&SFR&environment}.  The merger history of a galaxy may also be encoded in its morphology \citep[e.g.,][]{Hopkins2009MergerDiskSurvival}.  

Galaxies exhibit considerable evolution in various physical properties and morphology over cosmic time \citep[e.g.,][]{Wuyts2011StructureEvolution, HuertasCompany2016}.  Understanding the relationship between these evolutionary trends is crucial for completing the picture of how galaxies develop, transform, and assemble their mass.  Morphological studies at different redshifts show significant evolution.  As the lookback time increases, disks decrease in size \citep{Cassata2013ETGs, Margalef-Bentabol2016, Sachdeva2019DiskFormation}, clumpy and/or irregular features become more common \citep{Griffiths1994EarlyHSTMorphologies, Abraham1996EarlyHSTMorphologies, Mortlock2013HubbleSequenceFormation, Guo2015ClumpsI, HuertasCompany2016}, and mergers are expected to be more frequent \citep[but see][]{Man2016Mergers, Mantha2018MergerRate, Duncan2019MergerRates}.  As the cosmic SFR density changes with time \citep{Madau&Dickinson2014}, so do the relative SFR contributions of different morphologies.  In (s)SFR$-M_*$ space, disky late-type systems dominate the star-forming main sequence (MS), while spheroidal early-type systems dominate the quiescent population, at all redshifts \citep{Wuyts2011StructureEvolution, Lee2013GalaxyStructure}.  This general picture is overly simplified, however, as spheroids are present on the MS and disks are not uncommon off of it \citep[e.g.,][]{Brennan2015Quenching&MorphTransform}.  The approach taken in this paper to study the link between morphology and SFR is by analyzing the contribution to the SFR budget for different morphological types, as well as characterizing the typical SFRs and range of SFRs for different types.  For this approach to be successful, one requires robust estimates of the SFR and a method for quantifying the morphology, and both must be consistent across redshifts; this is the approach our study aims to produce.  Characterization of SFRs in relative terms, i.e., compared to what is typical at that redshift, is especially informative.  

Many methods have been introduced to quantify galaxy morphology, each with a different set of strengths and limitations.  Automated methods have been developed to more efficiently classify large samples and to quantify morphological features which are difficult or impossible to estimate by eye.  Parametric methods like the Sersic index and bulge-disk decomposition \citep{Sersic1963Profile, Freeman1970DisksofSpirals&S0s, Peng2002GALFIT} have enabled the study of bulge buildup and its effect on the quenching of star formation \citep[e.g.,][]{Kormendy2009Spheroids, Brennan2015Quenching&MorphTransform}.  Non-parametric statistics like the Gini coefficient, \textit{M}\textsubscript{20}, multiplicity ($\Psi$), \textit{CAS}, and \textit{MID} \citep{Conselice2003CAS, Abraham2003GiniCoefficient, LotzPrimackMadau2004GiniM20, Law2012Multiplicity, Freeman2013MID} are able to capture more complex or amorphous characteristics of the light distribution, making them especially useful for identifying disturbed morphologies \citep{Conselice2014GalaxyStructure&EvolutionReview}.  These methods have proven invaluable in investigating the statistical properties of large samples of galaxies at different redshifts \citep[e.g.][]{Mendez2011AEGISGreenValley, Wuyts2011StructureEvolution, Lee2013GalaxyStructure}.  

Despite their success, there are limits to the effectiveness of automated methods.  Information about the full light distribution is inevitably lost when using simple model-derived parameters or statistics.  This has led to the application of machine learning to facilitate visual-like classifications in an automated way \citep{HuertasCompany2015CANDELSNeuralNet, Dominguez2018DeepLearningGZoo2Catalog, Hocking2018UnsupervisedML}.  Though machine learning is quite effective and continues to see improvement, the level of detail that can be extracted is still somewhat limited.  A set of galaxies which have already been classified by eye is also sometimes required to train the machine learner. 

Unlike automated methods, human classifiers are able to visually process and interpret the full complexity of the light distribution in a galaxy image.  Visual morphologies remain contentious, however, because the resulting classifications can be subjective and influenced by biases introduced by differences in apparent size, surface brightness, and signal to noise ratio (among other factors).  Despite these drawbacks, direct visual classification remains a valuable and straightforward method for samples of modest size.  Visual classification is especially useful for identifying complex morphological features, which are difficult for automated schemes to identify, including disk substructures such as rings, bars, and spiral arms, as well as merger signatures such as tidal tails, loops, or bridges. % Maybe give some useful references here...?

Many morphological catalogs based on visual classification are available, especially for galaxies in the local universe.  The Galaxy Zoo project is famous for its crowd-sourcing approach, making use of public volunteers to visually classify thousands of galaxies in archival SDSS and \textit{HST} images \citep{Lintott2008GalaxyZoo1, Willett2017GZooHubble, Simmons2017GZooCANDELS}.  Expert classifications are also available, some with relatively large sample sizes \citep[i.e.,][]{Corwin1994RC3Corrected, Fukugita2007VisualCatalog, Nair&Abraham2010LocalVisualCatalog}.  A catalog of expert classifications for high-redshift ($0.3 \lesssim z \lesssim 3$) galaxies is provided by \citet{K15} (hereafter K15), who used high-resolution \textit{HST} images and a team of 65 classifiers to form an extensive accounting of structure and morphology in the portion of the GOODS-S field covered by CANDELS.  The \citetalias{K15} visual classifications have been used for calibration of automated classification methods \citep{HuertasCompany2015CANDELSNeuralNet, Peth2016PCAonK15, PerezCarrasco2019K15TrainedCLASHNeuralNet} and studies of merging or interacting systems out to high redshifts \citep{Silva2018CANDELSMergers, Pearson2019Mergers&SFRwMachineLearning}.  

Although there exist large samples of morphologically classified galaxies at different redshifts, and many studies that make use of them, there remain a number of challenges.  The assessment of the morphology of a galaxy depends on the resolved physical scale and rest-frame wavelength coverage of the images used for classification.  Furthermore, differences in classification schemes and methodology make it difficult, if not impossible, to reliably compare results from different studies.  This is made worse by the varying sensitivity of common morphological indicators to specific features, e.g. due to mass-to-light ratio effects \citep[see][]{Tacchella2015MorphKinematics&ColorsHST}.  In this work, we use a simplified classification scheme based on \citetalias{K15} for our high-redshift galaxies and apply the same simplified \citetalias{K15} scheme to lower redshifts, using images that are matched in both rest-frame wavelength coverage and physical resolution scale to the images used in \citetalias{K15}, creating a sample of consistently classified galaxies spanning a wide range of redshifts.  

Understanding the link between morphology and star formation also requires consistent and reliable SFR estimates.  SFRs can be derived using SED fitting, which involves fitting models of synthesized galaxy spectra to galaxies' observed broadband photometry.  SED fitting is a flexible and powerful tool which allows specification of various parameters including stellar evolutionary models, star formation histories, dust attenuation, and metallicity \citep[for a review, see][]{Conroy2013SEDFittingReview}.  Despite its utility, SED fitting is subject to systematic effects arising from uncertainties in the assumed models.  Estimates of the SFR are especially sensitive to the assumptions regarding the dust attenuation curve \citep{Salim2020GalaxyDustAttReview}.  The use of free dust attenuation curves has been shown to produce SFR estimates that are less biased than those derived using a universal curve \citep{Kriek&Conroy2013DustLawSystematics, Salim2016GSWLC, Salim2018DustAttCurves}.  An alternative method to derive SFRs by summing up UV and IR luminosites, which circumvents dust attenuation curve assumptions, is limited by relatively low completeness of IR data at high redshift and their potential contamination by AGN \citep{Daddi2007MidIRExcess}.  In this work, we derive SFRs for the entire sample using a consistent SED fitting process with a variable dust attenuation law.  

We describe our data and sample selection methods in Section \ref{Section:Data&SS}, while our morphological classifications and methodology are elucidated in Section \ref{Section:MorphologicalClassification}.  We then present our analysis and results in Section \ref{Section:Results}, discuss their implications in Section \ref{Section:Discussion}, and summarize our conclusions in Section \ref{Section:Conclusions}.  We include additional details regarding the degradation of local galaxy images in Appendix \ref{Appendix:ImageDeg}, present the \citetalias{K15} catalogs and our conversion to the simplified classifications in Appendix \ref{Appendix:K15Catalogs}, and provide a discussion of visual classification biases in Appendix \ref{Section:BiasCheck}.  Unless otherwise stated, we use AB magnitudes and a flat WMAP7 cosmology ($H_0 = 70$ km/s/Mpc, $\Omega_m = 0.27$).  

\section{Data $\&$ Sample Selection} \label{Section:Data&SS}

\begin{figure*}
    \centering
    \includegraphics[scale=0.45]{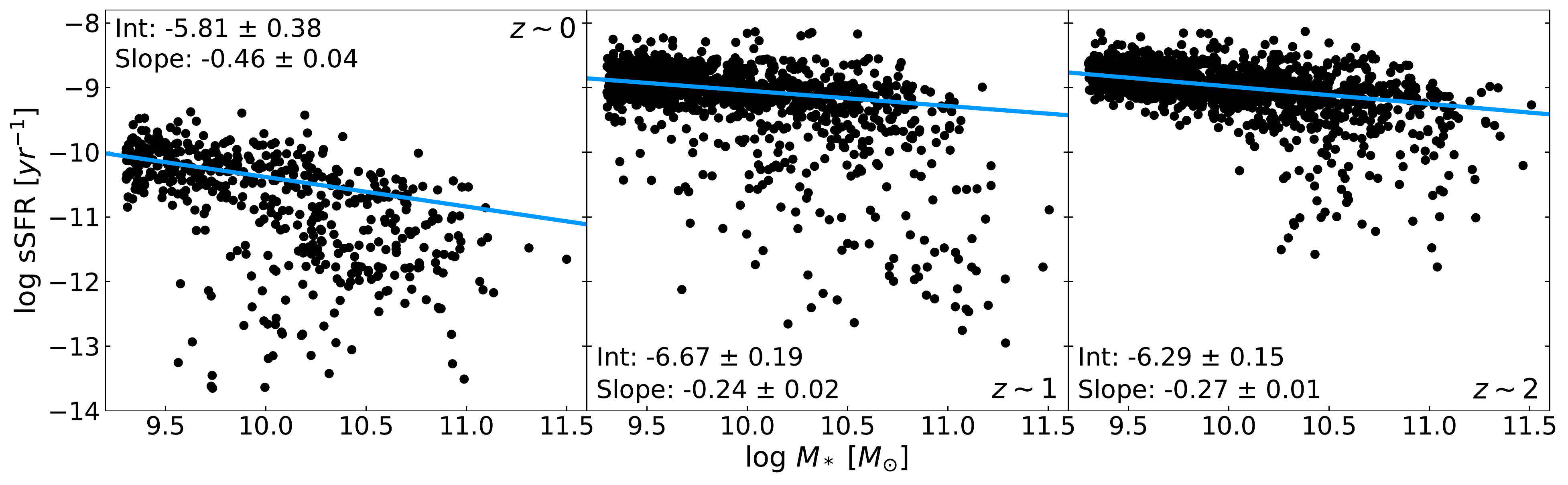}
    \caption{sSFR-$M_*$ diagrams for each of our three final redshift samples.  The low-redshift ($0.01 < z < 0.0176$), intermediate-redshift ($0.8 < z < 1.4$), and high-redshift ($1.5 < z < 2.5$) samples contain 506, 1152, and 1438 galaxies, respectively.  Physical properties (including stellar masses and SFRs) are derived using a consistent SED fitting procedure with flexible attenuation laws.  Our samples are complete above log$(M_*/M_{\odot}) \geq 9.3$ at $z \sim 0$ and $z \sim 1$, and above log$(M_*/M_{\odot}) \geq 10$ at $z \sim 2$.  However, the $\sim40\%$ incompleteness at $z\sim2$ in the mass range $9.3 < $ log$(M_*/M_{\odot}) < 10$ is not (s)SFR-dependent.  Blue lines represent the fits to the star-forming main sequence, as described in Section \ref{Section:Results:sSFRDistributions}.  The slopes and intercepts of these fits are shown in each panel.  The vertical offset with respect to these fits ($\Delta$ log sSFR$=-1$) is used to define our star-forming galaxy (SFG) and `red sequence' (RS) samples.}
    \label{SSFRvM_All}
\end{figure*}

In this section, we describe the selection process for each of our three redshift samples, the completeness of each sample, and the derivation of the physical parameters used in our study.  

\subsection{Intermediate and High-Redshift Samples} \label{dataGOODSS}

We use data from GOODS-S, which is one of the five principal CANDELS fields and covers $\sim 170$ arcmin\textsuperscript{2} of sky.  The photometric catalog of \citet[][hereafter G13]{Guo2013PhotoCat} combines UV to Mid-IR observations in GOODS-S from various public datasets, including CANDELS \citep{Grogin2011CANDELS, Koekemoer2011CANDELS} and the \textit{HST}/WFC3 Early Release Science (ERS) \citep{Windhorst2011ERS}.  The \citetalias{Guo2013PhotoCat} source detection was performed with SExtractor on the \textit{HST}/WFC3 F160W band imaging, in an alternated `hot' and `cold' mode designed to identify both very bright and very faint sources.  

The extensive wavelength coverage of the catalog allows for a detailed modeling of the spectral energy distributions (SEDs) of our galaxies across the rest-frame UV and optical ranges.  We select our high ($z \sim 2$) and intermediate ($z \sim 1$) redshift samples from \citetalias{Guo2013PhotoCat} based on the `best' redshift from \citet{Santini2015GOODSSMassCatalog}, which is either the photometric redshift or the spectroscopic redshift when the latter is available\footnote{Spectroscopic redshifts are available for $<10\%$ of objects.}.  For $z \sim 2$ we select objects in the $1.5 <$ \textit{z}\textsubscript{best} $< 2.5$ window and for $z \sim 1$ in $0.8 <$ \textit{z}\textsubscript{best} $< 1.4$, giving us a preliminary sample of 9907 galaxies at $z \sim 2$ and 6648 at $z \sim 1$.  The physical resolution of the images (in kpc/\arcsec) changes by a factor of only $\approx 5\%$ between $z = 1$ and $z = 2$, ensuring that the visual classification is not affected by resolution systematics.    

To assign visual morphologies to our GOODS-S sample, we first match to the \citetalias{K15} catalog of visual classifications.  The \citetalias{Guo2013PhotoCat} and \citetalias{K15} catalogs are based on the same imaging data, but lack common object IDs.  Based on the RA and DEC residuals of a $1\arcsec$ test matching, we find that a matching radius of $0.1\arcsec$ is sufficient to exclude spurious matches.  

The application of an \textit{H}-band magnitude cut of 24.5 in \citetalias{K15} results in a loss of low-mass galaxies.  Because the completeness of our sample does not depend on sSFR above log$(M_*/M_{\odot}) \sim 9.3$ at $1.5 < z < 2.5$, we adopt log$(M_*/M_{\odot}) = 9.3$ as our lower mass limit.  We apply the mass cut to the matched sample and arrive at the final sample sizes of 1438 and 1152 at $z \sim 2$ and $z \sim 1$, respectively.  We discuss the completeness of these samples in Section \ref{Data&SS:Completeness}.  The sSFR-$M_*$ distributions of the final samples are shown in Figure \ref{SSFRvM_All}.  

\subsection{Completeness of the Intermediate and High Redshift Samples}
\label{Data&SS:Completeness}

\begin{figure}
    \centering
    \includegraphics[scale=0.35]{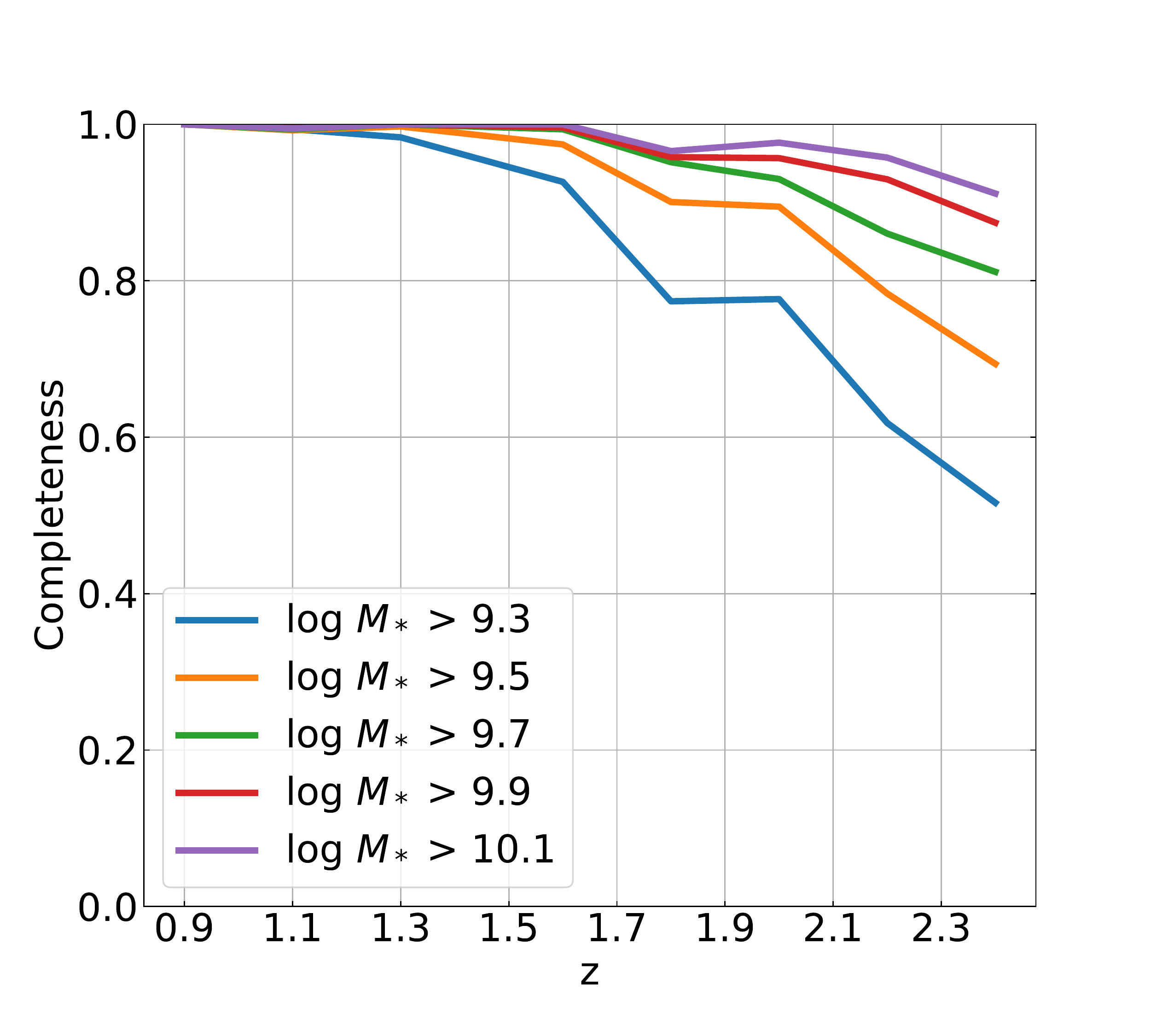}
    \caption{Completeness of our GOODS-S sample at different redshifts and mass cuts after applying the \citetalias{K15} \textit{H}-band magnitude cut ($H < 24.5$).  Our sample is essentially complete for high masses (log $(M_*/M_{\odot}) > 10$) at all redshifts.  At intermediate redshifts ($0.8 < z < 1.4$), the completeness is nearly unity at all masses.  The magnitude cut results in a loss of $\approx 38\%$ of our low-mass ($9.3 <$ log$(M_*/M_{\odot}) < 10$) galaxies in $z\sim 2$ bin.}
    \label{Completenessvsredshift}
\end{figure}

One potential source of incompleteness for our GOODS-S samples arises due to the source detection efficiency of the SExtractor setup used in \citetalias{Guo2013PhotoCat}.  This is especially significant for sources with apparent magnitudes close to the limiting depth of the survey.  To quantify this incompleteness, \citet{Duncan2014SMFCANDELS} ran a synthetic catalog of thousands of mock galaxies through the same SExtractor procedure used in \citetalias{Guo2013PhotoCat}, determining the completeness as a function of apparent \textit{H}-band magnitude (see Figure 3 of \citealt{Duncan2014SMFCANDELS} for details, and also \citealt{Duncan2019MergerRates}).  For any region in the field, the completeness at apparent \textit{H}-band magnitudes brighter than 24.5 is nearly unity.  As the \citetalias{K15} sample was subject to an \textit{H}-band magnitude cut of $<$ 24.5 mag, we can safely ignore the source detection incompleteness.  

Our second completeness check arises from requiring a match in \citetalias{K15}.  Matching between our greater GOODS-S catalog and the \citetalias{K15} catalog, $\approx 16\%$ of the GOODS-S galaxies brighter than the nominal \citetalias{K15} magnitude cut do not have a match in the \citetalias{K15} catalog.  These galaxies are located around the edges of the field and are largely separated from the matched galaxies; they were apparently excluded from \citetalias{K15} due to the depth and image quality concerns.  Because the region occupied by these cut galaxies has little to no overlap with the region containing our sample, they have no bearing on our analysis.  

To determine the volume completeness of our GOODS-S samples (prior to the \citetalias{K15} matching), we use \textit{H}-band magnitude limits from the \citetalias{Guo2013PhotoCat} catalog to calculate limiting redshifts (i.e., the redshift at which the galaxy becomes too faint to be included in our sample) for each galaxy in our $z \sim 2$ sample.  We find that none of the galaxies above our log$(M_*/M_{\odot}) = 9.3$ mass cut have limiting redshifts lower than the upper bound of our sample ($z < 2.5$), even after applying an empirical k-correction.  This suggests that our sample suffers no incompleteness resulting from the magnitude limits of the GOODS-S field.  

The final source of incompleteness in our sample is the \textit{H}-band magnitude cut ($H < 24.5$) imposed by \citetalias{K15}.  We determine this incompleteness using magnitudes from \citet{vanderWel2012CANDELSStructuralParams} and masses from our SED fitting (Section \ref{MethodsSED}).  We show the completeness as function of redshift for different mass cuts in Figure \ref{Completenessvsredshift}.  The sample is effectively complete for log$(M_*/M_{\odot}) > 10$ at all redshifts.  There is also almost no incompleteness in our $z \sim 1$ sample.  Galaxies lost due to the magnitude cut are largely limited to low masses close to the redshift limit (i.e., $2.0 < z < 2.5$).  At $z \sim 2$ the magnitude cut results in a loss of $\approx 38\%$ of our low-mass $9.3 <$ log$(M_*/M_{\odot}) < 10$ galaxies.  However, the completeness is not dependent on (s)SFR, preserving the relative composition of the MS at each mass.  Since our analysis is primarily relative (offset with respect to the MS; SFR and number fractions), the incompleteness is not a significant issue.  Notably, employing a stricter mass cut (log$(M_*/M_{\odot}) > 9.7$) does not meaningfully affect our main conclusions.

\subsection{Low-Redshift Sample} \label{DataLowz}

\begin{figure*}[ht]
\centering
\includegraphics[scale=0.9]{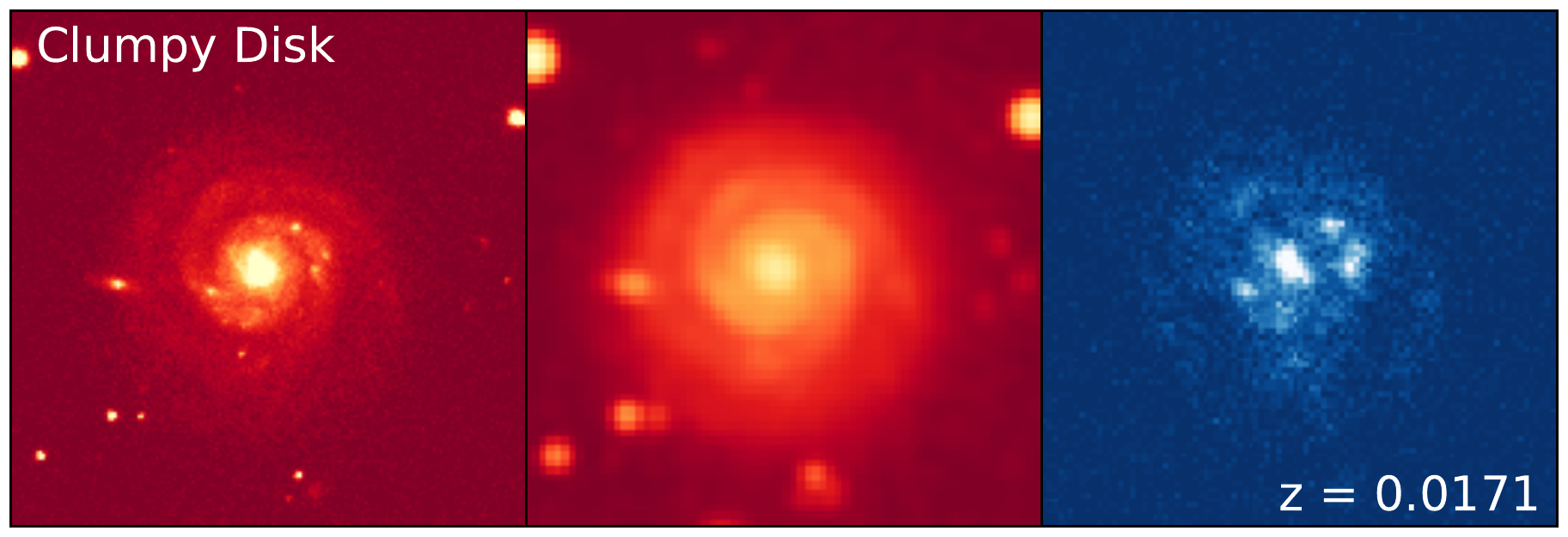}
\caption{An example of a $z \sim 0$ Clumpy Disk illustrating the effect that the image degradation (described briefly Section \ref{DataLowz}, in detail in Appendix \ref{Appendix:ImageDeg}) has on the visual appearance of SDSS galaxies.  We show the original SDSS \textit{r}-band image (left), the degraded \textit{r}-band image (middle), and the GALEX FUV image (right).  The degraded \textit{r}-band image has a physical resolution comparable to the \textit{HST} \textit{H}-band (F160W) images shown in Figure \ref{ClassImages}, whereas the FUV image, without any degradation, corresponds to the physical resolution of the \textit{HST} \textit{V}-band (F606W) images.}
\label{LocalConvExample}
\end{figure*}

Our low-redshift ($z \sim 0$) galaxies are drawn from the \textit{GALEX}-SDSS-WISE Legacy Catalog 2\footnote{GSWLC-2 is available at:  \url{http://pages.iu.edu/~salims/gswlc/}} \citep[GSWLC-2;][]{Salim2016GSWLC, Salim2018DustAttCurves}, which provides stellar masses, SFRs, and redshifts for $\sim 700,000$ SDSS spectroscopic galaxies at $z < 0.3$ with \textit{GALEX} UV coverage.  For our initial selection we use the X2 catalog, which includes SED fitting parameters based on the deepest UV imaging available for each object (the exposure time ranges from very shallow to very deep).  To have a matching physical resolution between \textit{GALEX} UV imaging and \textit{HST} rest-frame UV imaging at $z \sim 1$ and $z \sim 2$ we need to select low-redshift galaxies as close to the low-redshift cutoff of GSWLC (z=0.01) as possible.  We select a sample of 506 log$(M_*/M_{\odot}) > 9.3$ galaxies in the redshift range $0.01 \leq z \leq 0.0176$, satisfying SED fit quality (SED flag = 0) and with medium or deep FUV imaging (UV flag = 1 or 3, UV survey = 2 or 3).  GSWLC is complete above the mass cut throughout the volume encompassed by this redshift range.  We show the sSFR-$M_*$ distributions for our $z \sim 0$, $z \sim 1$, and $z \sim 2$ samples in Figure \ref{SSFRvM_All}. 

We use FUV images from \textit{GALEX} \citep{Morrissey2007GALEX, Bianchi2017GALEX} as well as monochromatic \textit{u}, \textit{g}, and \textit{r} images from SDSS DR9 \citep{SDSSDR9} to classify our low-redshift sample.  Image cutouts for each galaxy are generated using the SkyView package in Astropy \citep{Astropy2013, Astropy2018}.  Our selection of galaxies around the low-redshift cutoff of GSWLC ensures that the GALEX FUV images have comparable rest-frame physical resolution to \textit{HST} at $z \sim 1$ and $z \sim 2$.  At such low redshifts, however, SDSS has a finer physical resolution than \textit{HST} in the rest-frame optical.  We therefore degrade our SDSS images prior to classification, using Astropy's \citep{Astropy2013, Astropy2018} Gaussian convolution and block reduction to smooth and resample the images, respectively.  Degrading the SDSS images ensures that our $z \sim 0$ galaxies are classified in a consistent manner to the \citetalias{K15}-matched sample.  The end result of the degradation process for one of the galaxies in our sample is shown in Figure \ref{LocalConvExample}.  We describe the image degradation in greater detail in Appendix \ref{Appendix:ImageDeg}.

\subsection{SED Fitting} \label{MethodsSED}

The masses, SFRs, and redshifts used in this study are derived using SED fitting with the CIGALE code \citep{Noll2009CIGALE, BoquienCIGALE2019}, which allows specification of the SF history, dust attenuation, and includes modeling of the emission lines.  CIGALE uses a Bayesian methodology to estimate the physical parameters, constructing a probability distribution function (PDF) whose mean gives the adopted value for the given parameter.  Stellar emission is modelled using \citet{BC03} stellar population synthesis.  A Chabrier IMF \citep{ChabrierIMF2003} and flat WMAP7 cosmology ($H_0 = 70$ km s\textsuperscript{-1} Mpc\textsuperscript{-1}, $\Omega_m = 0.27$) are assumed.  The principal difference between the galaxy parameters used in this work compared to those available from the literature lies in the use of free dust attenuation curves.

A detailed description of the SED fitting method we employ for the low-redshift sample can be found in \citet{Salim2016GSWLC, Salim2018DustAttCurves}.  Photometry is primarily taken from \textit{GALEX} (FUV and NUV) and SDSS (\textit{ugriz}).  WISE mid-IR (12 and 22 $\mu$m) data are incorporated via their constraints on the IR luminosity.  The derived dust attenuation curves span a range of values and are on average significantly steeper than the \citet{Calzetti2000SFGDust} curve, which results in systematically different estimates of parameters, in particular the SFR.  

We use the same SED fitting code and similar model libraries (but adjusted for young galaxies) to derive the physical parameters of the intermediate and high-redshift galaxies.  Photometry is taken from \citetalias{Guo2013PhotoCat} and includes data from Blanco (\textit{U}), VLT/VIMOS (\textit{U}), \textit{HST}/ACS (F435W, F606W, F775W, F814W, F850LP), \textit{HST}/WFC3 (F125W, F160W), VLT/ISAAC (\textit{Ks}), VLT/HAWK (\textit{K}), and Spitzer/IRAC (3.6, 4.5, 5.8, and 8 $\mu$m).  The SED fitting for GOODS-S does not use constraints from dust emission (IR luminosity) because of the potential contamination by AGN emission, especially at $z \sim 2$ \citep{Daddi2007MidIRExcess}, but based on our own analysis of $z \lesssim 1$ galaxies we find that its absence does not lead to systematic differences in SFR or stellar mass, only less accurate quantities.  In contrast, the inclusion of IR data when fitting $z \sim 2$ galaxies leads to systematically elevated SFRs, and we have reason to believe that this is due to biases in the IR data.  

For the $z \sim 1$ and $z \sim 2$ samples we also recreate all of our main results (i.e., Figures \ref{deltaSSFRGauss} through \ref{RSFracEvol}) using the stellar masses and SFRs from \citet{Fang2018UVJCANDELS}, who use a combination of SED fitting results obtained with a fixed dust attenuation curve and explicit SFR calibrations.  These alternative results broadly agree with the ones derived with our nominal parameters.  The differences inform us of the trends that may not be statistically robust, and we note these differences where they are significant.  There do exist systematic differences in the sSFRs at $z \sim 2$, but since all of the analyses are relative (sSFR with respect to the MS; SFR contribution) this does not affect the results.  The comparison with \citet{Fang2018UVJCANDELS} as well as the effects of including IR data in SED fitting for $z \sim 2$ galaxies will be dealt with in more detail in a future paper.

\section{Morphological Classification} \label{Section:MorphologicalClassification}

In this section, we describe the morphological classification scheme used to assign the galaxies in our sample into mutually exclusive classes.  Additional details regarding the \citetalias{K15} data release and our conversion to a simplified scheme are provided in Appendix \ref{Appendix:K15Catalogs}.  We also investigate the impact of visual classification biases on our sample in Appendix \ref{Section:BiasCheck}.

\subsection{Intermediate and High-Redshift Samples} \label{GOODSS_Class}

\begin{figure*}[ht]
\centering
\includegraphics[scale=0.7]{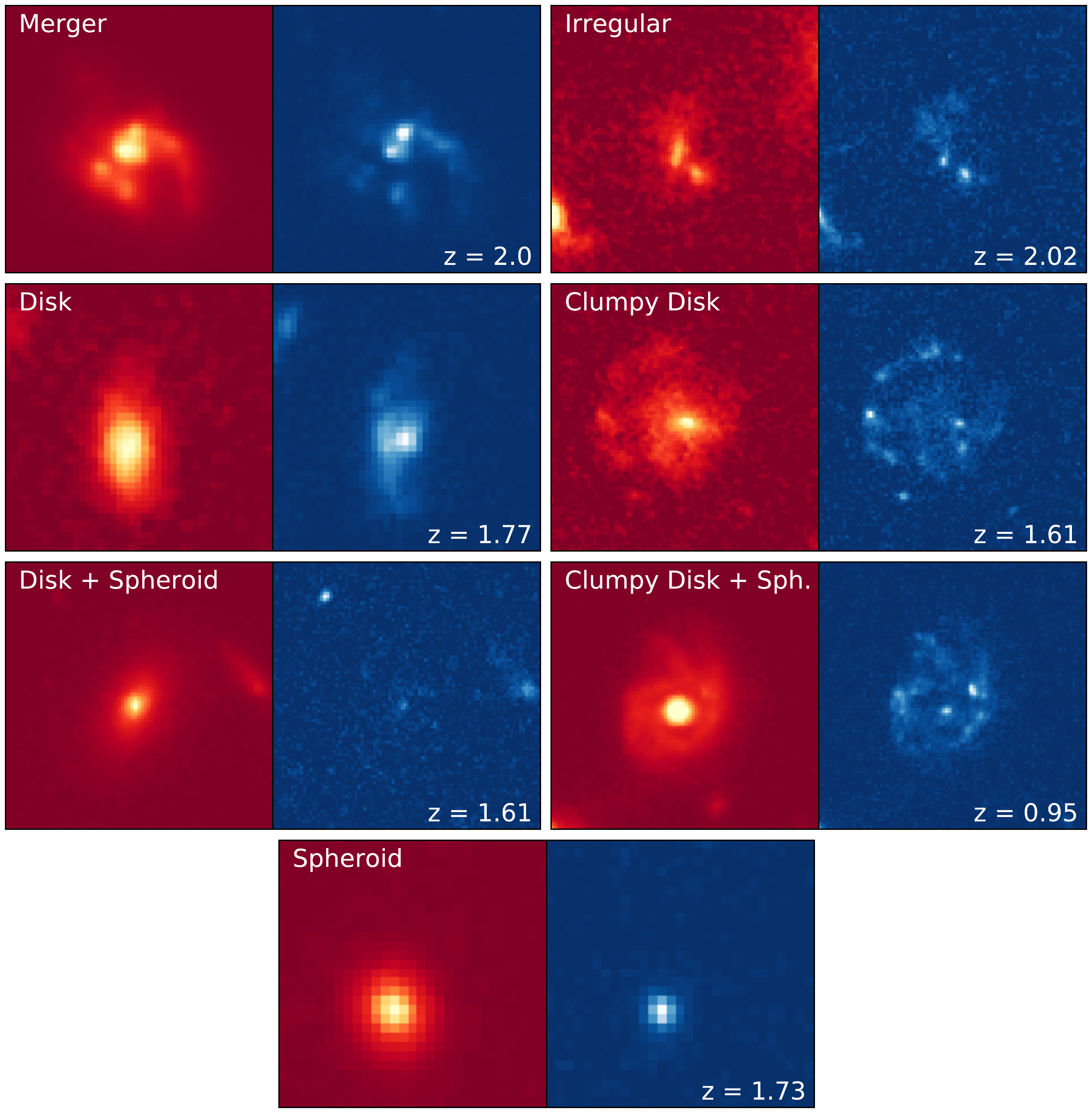}
\caption{A selection of \textit{HST} cutouts of GOODS-S galaxies from different morphological classes.  The \textit{H}-band (F160W) image is shown in red on the left whereas the \textit{V}-band (F606W) image is shown in blue on the right.  The size of each cutout is 8 times the circular \textit{H}-band half-light radius, taken from \citet{vanderWel2012CANDELSStructuralParams}. The image scaling has been adjusted for each galaxy to highlight their morphological features.  The Spheroid is slightly off-center due to an error in its reported RA/DEC from CANDELS.}
\label{ClassImages}
\end{figure*}

Visual classification for our high and intermediate redshift samples is based on the \citetalias{K15} catalog, which contains detailed morphological information for 7,634 galaxies in the GOODS-S field.  The images used for these classifications are from both the CANDELS and ERS surveys.  For a detailed classification scheme description, see Figure 3 of \citetalias{K15}.  Classifiers were shown an image of the object in four different filters:  F160W, F125W, F850LP, and F606W (corresponding to \textit{H}, \textit{J}, \textit{z}, and \textit{V}, respectively).  The classifications were based primarily on the \textit{H}-band image, corresponding to the rest-frame optical at $z \sim 2$.  The other three images served to inform the \textit{H}-band classification and allowed the identification of clumps and patches in the rest-frame UV (using mostly \textit{V}-band).  \citetalias{K15} defines clumps as concentrated, independent, off-center knots of light, whereas patches are more diffuse.  Classifiers were shown square cutouts scaled to the size of the galaxy (in the \textit{H}-band), and they were allowed to freely adjust the intensity scaling of each image.  A second, larger \textit{H}-band cutout was also shown to allow the interaction status to be better determined.  

The \citetalias{K15} classifications are not discrete; classifiers were allowed and encouraged to place objects in multiple categories wherever appropriate.  The number of classifiers per object ranged from 3 to 7, with the exception of the calibration set of 200 galaxies (used to test the classification scheme; see Section 4.1 in \citetalias{K15}) given to all 65 classifiers.  While K15 does not provide a classification into mutually exclusive classes, such a classification scheme can be derived by combining various features of the \citetalias{K15} catalogs.  We classify the sample into 9 mutually exclusive classes:  Disk, Clumpy Disk, Disk + Spheroid, Clumpy Disk + Spheroid, Spheroid, Irregular, Merger, Point Source / Compact (PS/C), and Unclassifiable.  Galaxies which do not satisfy the selection criteria for any of these classes are Uncategorized.  

Construction of the scheme has been informed by visual inspection of a large number of galaxies and consultation of their K15 morphological assignments.  To visually inspect our sample, we make use of F606W, F850LP, F125W, and F160W imaging data from the CANDELS v1.0 data release for the bulk of the field.  For the region of GOODS-S covered by the ERS, we use F125W and F160W images from the Hubble Legacy Fields \citep[HLF,][]{Illingworth2016HLF}.  Cutouts are chosen to be of fixed angular size and large compared to our galaxies ($\sim 5\arcsec \times 5\arcsec$) to better assess the classifications, especially in relation to the interaction status.  As in \citetalias{K15}, we allow free adjustment of the image scaling during the visual inspection.  We find that clumps visible in the rest-frame optical (\textit{H}, \textit{J}) are quite rare; we therefore use only the rest-frame UV images to identify clumps in our low-redshift sample.  We also visually check each of our Mergers and find the classifications to be reliable overall.  

Here we broadly summarize the key features that separate these classes from one another.  For more details on the \citetalias{K15} data release and our conversion to the simplified scheme, see Appendix \ref{Appendix:K15Catalogs}.

\noindent \textbf{1) Merger:}  Mergers are single galaxies which show evidence of tidal features such as tails or loops, highly irregular outer isophotes, or double nuclei.  These features are signposts for interactions that resulted in the coalescence of two galaxies.  \newline

\noindent \textbf{2) Disk:}  These galaxies possess disk structures which may have some features or irregularities.  They have relatively low central concentration and may feature spiral arms, though this is not a requirement.  \newline

\noindent \textbf{3) Clumpy Disk:}  Clumpy Disks are Disk galaxies with prominent clumps.  Clumps are concentrated, independent nodes of light appearing primarily in the rest-frame UV images.  \newline  
    
\noindent \textbf{4) Spheroid:}  Spheroidal galaxies are roughly round or ellipsoidal and/or possess high central concentration.  Highly irregular spheroids are placed into the Irregular class.  \newline
    
\noindent \textbf{5) Disk + Spheroid:}  Galaxies with majority votes for both spheroid and disk are placed in the Disk + Spheroid class.  These galaxies possess disks which are smoother and more centrally concentrated than pure disks.  Disks with prominent bulges fall within this class, such as the lenticular (S0) galaxies in the Hubble classification.  \newline

\noindent \textbf{6) Clumpy Disk + Spheroid:}  These are simply Disk + Spheroid galaxies that contain clumps. \newline
    
\noindent \textbf{7) Irregular:}  This class contains galaxies which are not readily classifiable into any other class due to their peculiar morphologies.  This could be induced by strong interactions (but not recognized as a Merger by the classifier) or be intrinsic to the galaxy itself. \newline

We show the \textit{H}-band and \textit{V}-band images for a selection of galaxies in our $z \sim 2$ sample from each class in Figure \ref{ClassImages}.  Galaxies in the PS/C, Unclassifiable, or Uncategorized classes are not shown.  The PS/C and Unclassifiable classes contain very few galaxies, together making up less than $3\%$ of our GOODS-S samples, and so have very little impact on the results.  Uncategorized galaxies have no defining characteristics save for poor classifier agreement, and make up no more than $7 \%$ ($3 \%$) of the sample at $z \sim 2$ ($z \sim 1$).  Uncategorized galaxies have similar mean properties (i.e., mass, magnitude) to the total sample; this suggests that they simply possess complex morphologies which are difficult to interpret, leading to disagreement among the classifiers and rendering such galaxies a poor fit for the simplified scheme of Figure \ref{DecisionTree}.  Notably, for $\sim 17\%$ of the Uncategorized galaxies the classifiers at least agree that the galaxy is interacting, thus interactions may also lead to classifier disagreement with respect to the main morphology class (i.e., disk, spheroid, or irregular).  We do not show results from the PS/C, Unclassifiable, or Uncategorized classes, though their number and SFR contributions are still factored in where applicable.  

The \citetalias{K15} classifications are well-correlated with the Sersic index (see Figure 12 of \citetalias{K15}).  We also compare our simplified scheme to the Sersic index in Figure \ref{SersicCompare} but divide each class by their relative star-formation activity, recovering the same general correlations (see Sections \ref{Section:Results} and \ref{Discussion:Spheroids} for more details).  Disks tend to have light profiles close to exponential ($n \sim 1$) while Spheroids tend towards higher Sersic indices (i.e., more centrally concentrated profiles).  \citetalias{K15} also find good correlation between their classifications and UVJ colors; Spheroids are abundant in the quiescent region while Disks and Irregulars are common in the star-forming region (see Figure 13 of \citetalias{K15}).  Classifications based on the \citetalias{K15} catalog therefore generally follow established relationships between galaxy morphology and intrinsic properties. 

\subsection{Low-Redshift Sample} \label{SDSSdeg}

To classify the low-redshift sample, we utilize monochromatic images from SDSS DR9 in the optical \citep{SDSSDR9} and \textit{GALEX} images in the UV \citep{Morrissey2007GALEX, Bianchi2017GALEX}.  The \textit{GALEX} FUV images serve the same role as the \textit{HST}/ACS \textit{V} and \textit{z} images, facilitating the identification of clumps.  For the optical regime we make use of SDSS \textit{u}, \textit{g}, and \textit{r} images, which together cover approximately the same rest-frame spectral range as the \textit{H} and \textit{J} bands at $z \sim 2$. 

We use a single classifier for our low-redshift sample.  Both Astropy \citep{Astropy2013, Astropy2018} and SAOImage DS9 \citep{Joye2003DS9} were used to facilitate the training process, which involved developing a familiarization with the definitions used in \citetalias{K15} as well as direct visual inspection of image cutouts for galaxies in our \citetalias{K15}-matched sample.  Galaxies were drawn and displayed randomly, from either the entire sample or from individual classes, in order to test the classifier.  

To classify our local sample we used the following process.  For each galaxy a DS9 window is called and four images are displayed.  Three of these are our degraded \textit{u}, \textit{g}, and \textit{r} images while the fourth is the unmodified \textit{GALEX} FUV image.  The image cutouts are initially chosen to have a fixed angular size of $\sim 300\arcsec \times 300\arcsec$ to ensure that the galaxy and any close companions are visible.  Galaxies are flagged and followed up with larger cutouts if they are exceptionally large in angular extent.  In keeping with the procedure outlined in \citetalias{K15}, the morphological classification is based primarily on the \textit{r}-band morphology, with the other three images informing the classification.  The classifier is allowed to freely adjust the image intensity scale but is not provided any information (e.g. mass, SFR, redshift) beyond the images themselves.  Only the FUV image is used to identify clumps.  To avoid classifier fatigue, classification is done in chunks of 50 or 100 galaxies over the course of a week.  After each round of classification, the chunk is reviewed and galaxies noted as ambiguous are double-checked.  

\section{Results} \label{Section:Results}

\begin{deluxetable*}{ccccccccc}[ht!]
\tablecaption{Number of galaxies of each morphological class in our sample and in each mass/redshift bin.  log $(M_*/M_{\odot}) = 10$ is our split between high and low mass.  The number fractions, relative to the total count in each bin, are given in parentheses and rounded to two decimal points.}
\tablecolumns{7}
\tablenum{1}
\label{SampleSizeTable}
\tablewidth{0pt}
\tablehead{
\colhead{Bin} &
\colhead{Disk} &
\colhead{Clumpy Disk} &
\colhead{Disk+Sph} & \colhead{Cl.Disk+Sph} & \colhead{Spheroid} & \colhead{Irregular} & \colhead{Merger} & \colhead{Total}
}
\startdata
$0.01 < z < 0.0176$ & & & & & & & & 506 \\
Low-mass SFG & 141 (0.62) & 21 (0.10) & 31 (0.14) & 1 (0.00) & 28 (0.12) & 3 (0.01) & 2 (0.01) & 229 \\
High-mass SFG & 69 (0.38) & 30 (0.16) & 59 (0.32) & 1 (0.01) & 16 (0.09) & 3 (0.02) & 4 (0.02) & 182 \\
RS & 2 (0.02) & 1 (0.01) & 52 (0.55) & 0 & 40 (0.42) & 0 & 0 & 95 \\
\hline
$0.8 < z < 1.4$ & & & & & & & & 1,152 \\
Low-mass SFG & 241 (0.36) & 152 (0.23) & 80 (0.12) & 29 (0.04) & 103 (0.15) & 29 (0.04) & 6 (0.01) & 670 \\
High-mass SFG & 92 (0.24) & 91 (0.24) & 65 (0.17) & 38 (0.10) & 63 (0.16) & 15 (0.04) & 3 (0.01) & 383 \\
RS & 12 (0.12) & 0 & 24 (0.24) & 0 & 57 (0.58) & 0 & 0 & 99 \\
\hline
$1.5 < z < 2.5$ & & & & & & & & 1,438 \\
Low-mass SFG & 279 (0.34) & 117 (0.14) & 77 (0.10) & 15 (0.02) & 152 (0.19) & 75 (0.10) & 24 (0.03) & 811 \\
High-mass SFG & 182 (0.31) & 96 (0.16) & 58 (0.10) & 13 (0.02) & 106 (0.18) & 51 (0.09) & 16 (0.03) & 586 \\
RS & 3 (0.07) & 0 & 6 (0.15) & 1 (0.02) & 25 (0.61) & 0 & 0 & 41 \\
\enddata
\end{deluxetable*}

\begin{figure*}[ht!]
    \centering
    \includegraphics[scale=0.225]{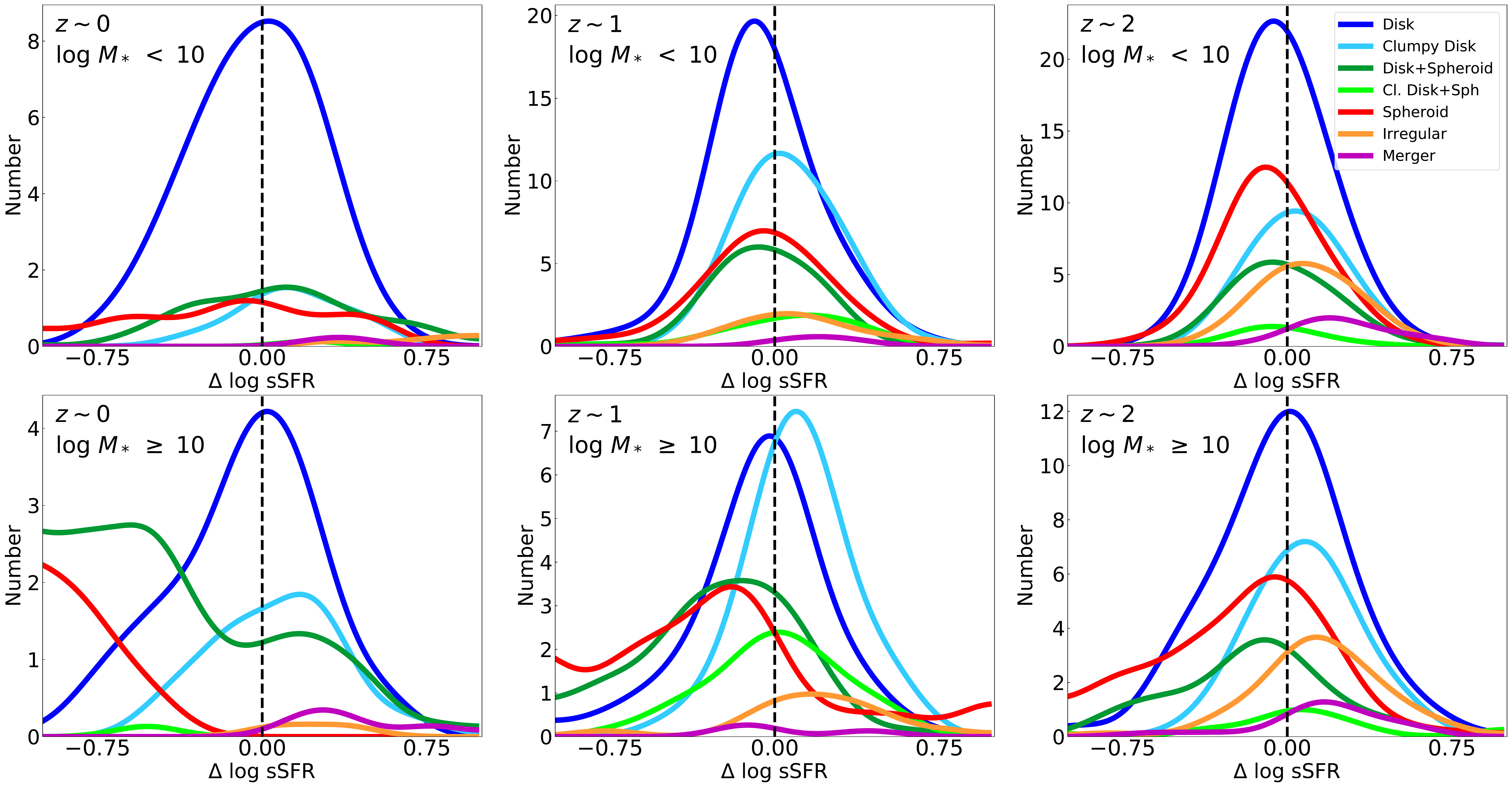}
    \caption{Smoothed number distributions in $\Delta$ log sSFR for galaxies of different morphological classes in two mass bins and three redshift bins.  The distributions shown here were formed by first binning the data in 0.05 dex bins, then applying a Gaussian kernel with $\sigma = 3$.  The position of the main sequence ($\Delta$ log sSFR $= 0$) is marked as a black dashed line in each panel.}
    \label{deltaSSFRGauss}
\end{figure*}

In this section, we describe the analysis performed on our morphologically-classified sample.  Section \ref{Section:Results:sSFRDistributions} describes our division into different SFR and mass bins and our method for characterizing the properties of star-forming galaxies (SFGs).  Section \ref{Section:Results:MSTypeFracs} discusses the number fractions for different classes of SFGs, while Section \ref{Section:Results:Mean&SigmaEvol} describes the evolution of the mean $\Delta$ log sSFR for different SFG classes.  We then present the SFR budget for SFGs in Section \ref{Section:Results:MSBudgetResults}, followed by the morphological composition of the low-sSFR `red sequence' (RS) in Section \ref{Section:Results:RedSequenceComposition}.  

\subsection{sSFR Distributions of Different Morphological Classes} \label{Section:Results:sSFRDistributions}

Our first goal is to determine where galaxies of different morphological classes are found with respect to the mean main sequence (MS) at $z \sim 0$ ($0.01 < z < 0.0176$), $z \sim 1$ ($0.8 < z < 1.4$), and $z \sim 2$ ($1.5 < z < 2.5$), and to establish the relative abundance of each class.  To parameterize the mean MS, we use linear least-squares regression to fit a line to the star-forming galaxies in sSFR-$M_*$ space at each redshift.  We fit only to galaxies (of all masses) above a fixed sSFR threshold in order to exclude quiescent outliers from the fitting.  We take the threshold to be log sSFR = -9.8 at $z \sim 1$ and $z \sim 2$, while at $z \sim 0$ we use log sSFR = -11.2.  The vertical offset from the regression line, $\Delta$ log sSFR, is then calculated for each galaxy, with a positive offset indicating an enhancement of (s)SFR relative to the MS.  We define our sample of star-forming galaxies (SFGs) to include all galaxies with $\Delta$ log sSFR $\geq -1.0$.  Our ``red sequence'' (RS) is then taken to be all galaxies with $\Delta$ log sSFR $< -1.0$.  Though we refer to galaxies below the MS collectively as the red sequence, this should be taken figuratively as we do not perform any color selection.  For most of the analysis, we do not distinguish between truly quiescent and transitional (green valley; \citealt{Salim2014GreenValley}) galaxies because of relatively small sample sizes and because of ambiguities in defining a green valley at intermediate and high redshifts.  We also mostly do not consider starbursts above the MS ($\Delta$ log sSFR $> 0.5$) because of very small sample sizes.  We split the SFGs into two mass bins, using log $(M_*/M_{\odot})$ = 10 as a dividing mass.  We show the number counts and fractions for each of our bins in Table \ref{SampleSizeTable}.  We show the MS fits, including the slopes and intercepts, in Figure \ref{SSFRvM_All}.  Errors in the slopes and intercepts are $\lesssim 10\%$ at all redshifts.  The equations for the MS line fits in each redshift bin are as follows:

$$ z \sim 0 : \quad \textrm{log sSFR} = -0.46 \times \textrm{log} (M_*/M_{\odot}) - 5.81 $$

$$ z \sim 1 : \quad \textrm{log sSFR} = -0.24 \times \textrm{log} (M_*/M_{\odot}) - 6.67 $$

$$ z \sim 2 : \quad \textrm{log sSFR} = -0.27 \times \textrm{log} (M_*/M_{\odot}) - 6.29 $$

To help visualize and compare the trends between different classes as a function of $\Delta$ log sSFR, we show in Figure \ref{deltaSSFRGauss} the smoothed $\Delta$ log sSFR distributions for galaxies of each morphological class in each redshift and mass bin.  The distributions are smoothed by first binning the galaxies in narrow 0.05 dex bins, then applying a Gaussian kernel with $\sigma = 3$ dex.  We show only the range $-1.0 < \Delta$ log sSFR $< 1.0$ to better emphasize the characteristics of the distributions in the star-forming region, which constitutes the bulk of our analysis.  We use the smoothed distributions to determine the number fractions of galaxies as a function of $\Delta$ log sSFR for each bin, which we show in Figure \ref{deltaSSFRFracs}.  The means of these distributions are used to create Figure \ref{deltaSSFRMeansvZ}.  

\subsection{Morphological Composition of the Star-Forming Galaxies} \label{Section:Results:MSTypeFracs}

\begin{figure*}
    \centering
    \includegraphics[scale=0.44]{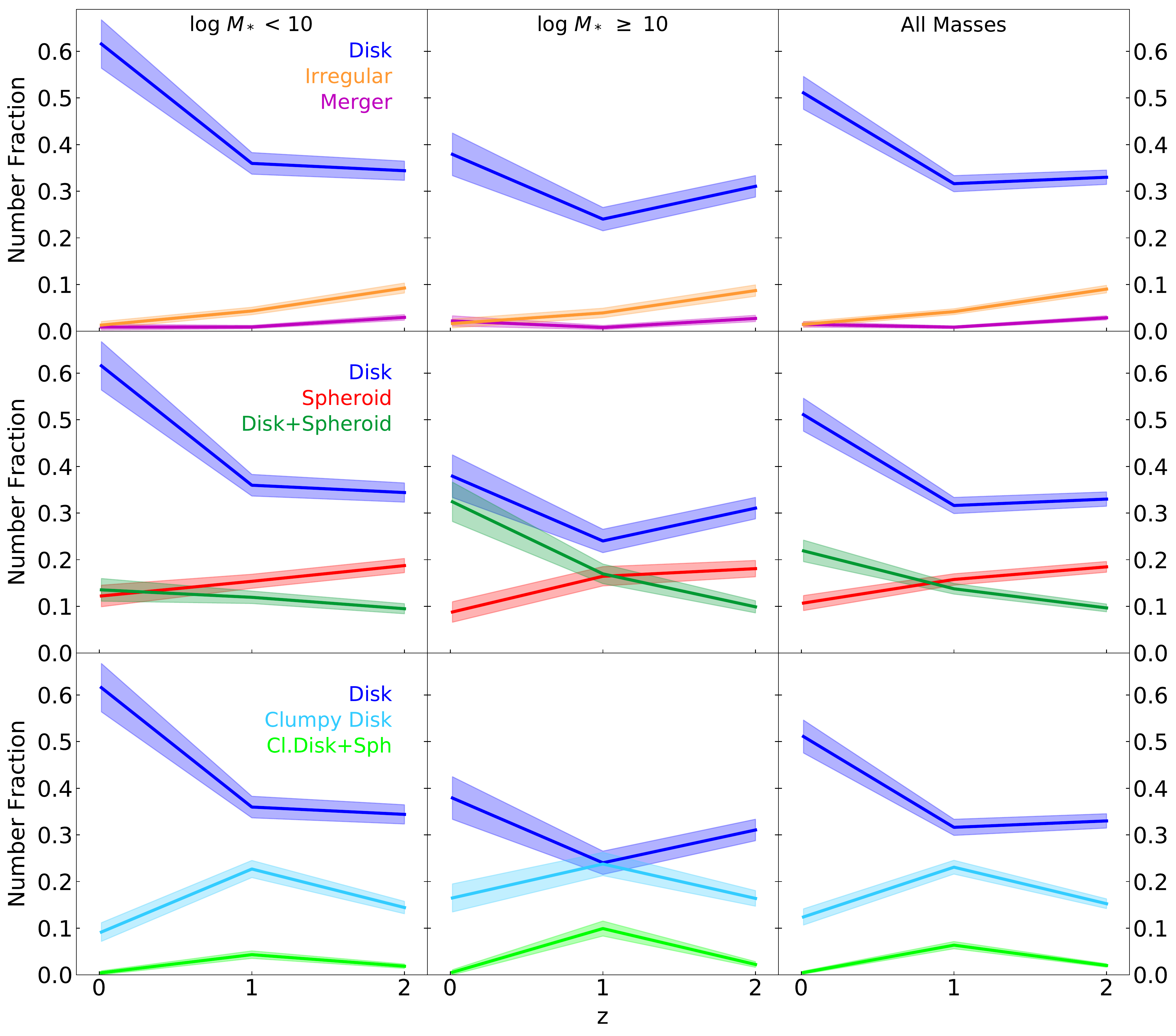}
    \caption{Evolution of the number fractions for star-forming galaxies (SFGs).  Disks form the bulk of SFGs at all masses and redshifts.  Clumpy disks with or without spheroids are more common at high mass and peak in contribution at $z \sim 1$.  Clumpy Disks remain common at $z \sim 0$, especially at high mass where their fraction matches that of $z \sim 2$.  Star-forming Spheroids are common ($\gtrsim 10\%$) at all redshifts, but have the lowest contribution at $z \sim 0$.  Mergers are rare at all redshifts.  The SFGs are more diverse at $z \sim 2$ where different classes contribute more equally.  Errors on the number fractions are the Poisson-like errors ($\sqrt{N}$).}
    \label{TypeFracEvol}
\end{figure*}

\begin{figure*}[ht]
    \centering
    \includegraphics[scale=0.215]{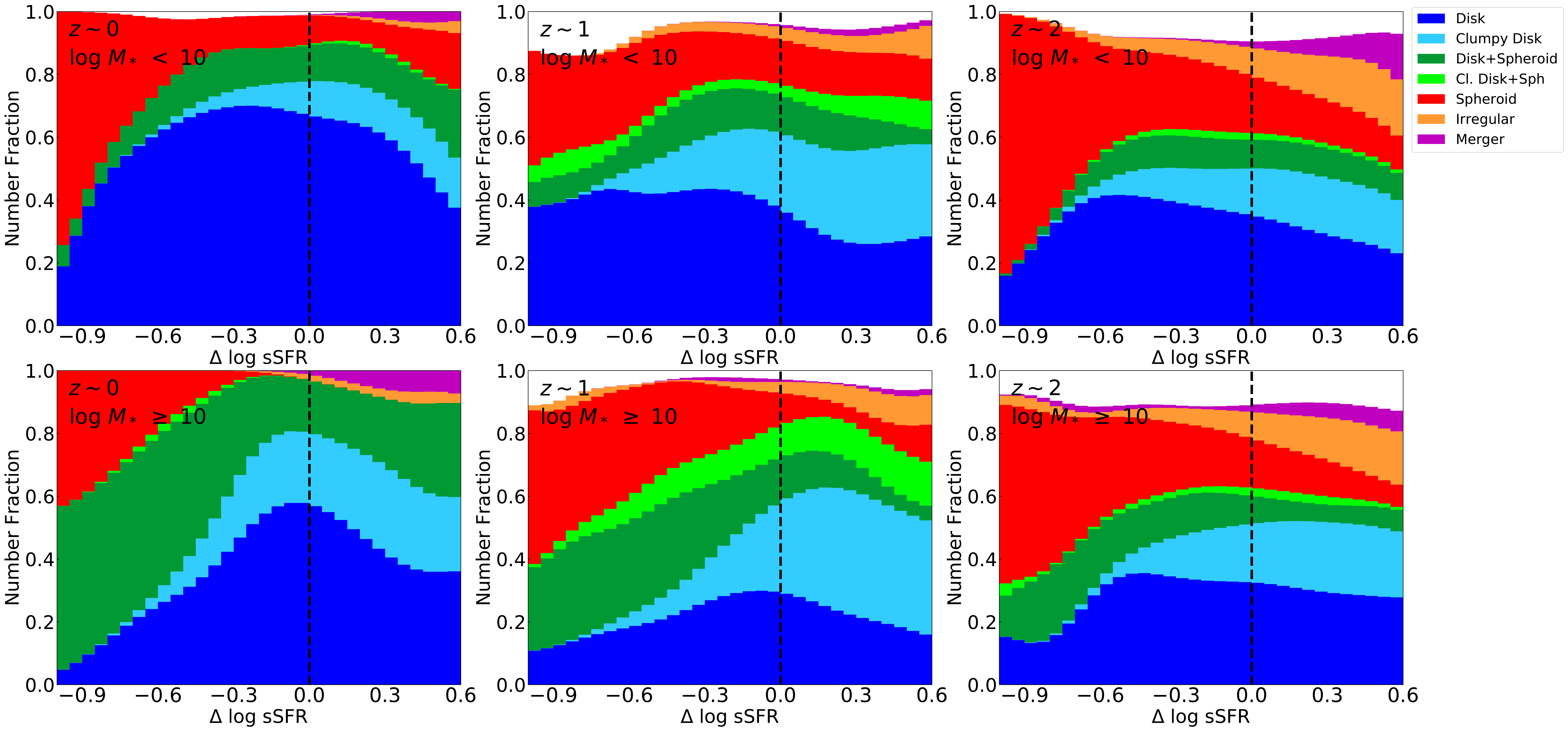}
    \caption{Stacked number fraction distributions vs. offset from the MS for different morphological classes.  Fractions are determined at each $\Delta$ log sSFR using the smoothed distributions of Figure \ref{deltaSSFRGauss}.  We include galaxies between $-1.0 < \Delta$ log sSFR $< 0.6$, within the star-forming region.  We exclude regions above 0.6 dex because the sample size becomes very low.  The position of the main sequence ($\Delta$ log sSFR $= 0$) is marked as a black dashed line in each panel.  The contribution of galaxies classified as either PS/C, Unclassifiable, or Uncategorized are not shown explicitly but are still included, hence the bars do not always add up to unity; this is most apparent at $z \sim 2$ where such galaxies are most common.  Star-forming Spheroids increase in number fraction with decreasing $\Delta$ log sSFR regardless of the mass or redshift.  Mergers tend to have higher fractions above the MS, but never more than $\sim 8\%$ of galaxies even in the starburst regime ($\gtrsim 0.5$ dex).}
    \label{deltaSSFRFracs}
\end{figure*}

In this section, we present the number fractions of different morphologies for low- and high-mass star-forming galaxies (SFGs; $\Delta$ log sSFR $\geq -1.0$).  We show the redshift evolution of these fractions in Figure \ref{TypeFracEvol}.  To help inform the results of Figure \ref{TypeFracEvol}, we also provide the number fraction as a function of $\Delta$ log sSFR for each mass and redshift bin in Figure \ref{deltaSSFRFracs}.  We first consider the low-mass results.  At low mass, Disks have the highest fractions at all redshifts and become dominant ($\sim 60\%$) at $z \sim 0$.  Clumpy Disks and Clumpy Disk + Spheroids both peak in fraction at $z \sim 1$, with the former remaining fairly common ($\sim 10\%$) even at $z \sim 0$.  Spheroids are common ($> 10\%$) at all redshifts, though their fractions decrease slightly over time.  Spheroid fractions also increase with decreasing $\Delta$ log sSFR at all masses and redshifts.  Irregulars are not uncommon at $z \sim 2$ where they form $\sim 10\%$ of the SFGs, but decrease in contribution steadily over time and become quite rare at $z \sim 0$.  Mergers are rare at all redshifts, never forming more than a few percent of the SFGs, but are most common at $z \sim 2$.  Mergers above the MS ($\Delta$ log sSFR $\gtrsim 0.5$) have elevated fractions compared to Mergers among all SFGs, but nonetheless maintain consistently low fractions ($\lesssim 8\%$) at all redshifts. 

Shifting our attention to high masses, the general picture is similar to that at low mass, though there are some crucial differences.  Disks are overall less dominant, only forming $\sim 40\%$ of the SFGs at $z \sim 0$.  The high-mass Disk fractions are actually lowest at $z \sim 1$ where the Clumpy Disk and Clumpy Disk + Spheroid fractions peak; this may suggest that many normal Disks at $z \sim 2$ will form clumps by $z \sim 1$.  Clumpy Disks have larger fractions than at low mass for all redshifts, even matching the Disk fraction at $z \sim 1$.  Interestingly, Clumpy Disks have similar fractions ($\sim 15\%$) at $z \sim 0$ and $z \sim 2$.  Clumpy Disk + Spheroids also have higher fractions, except at $z \sim 0$ where they remain effectively absent.  Notably, Disk + Spheroids are more abundant than at low mass for $z \lesssim 1$, and increase in fraction from nearly $20\%$ at $z \sim 1$ to over $30\%$ at $z \sim 0$.  Mergers show more erratic behavior, decreasing in fraction from $z \sim 2$ to $z \sim 1$ but then increasing from $z \sim 1$ to $z \sim 0$.  This may be attributable to the small sample sizes involved (see Table \ref{SampleSizeTable}).

\subsection{Evolution of the Relative sSFRs}
\label{Section:Results:Mean&SigmaEvol}

\begin{figure*}[h]
    \centering
    \includegraphics[scale=0.425]{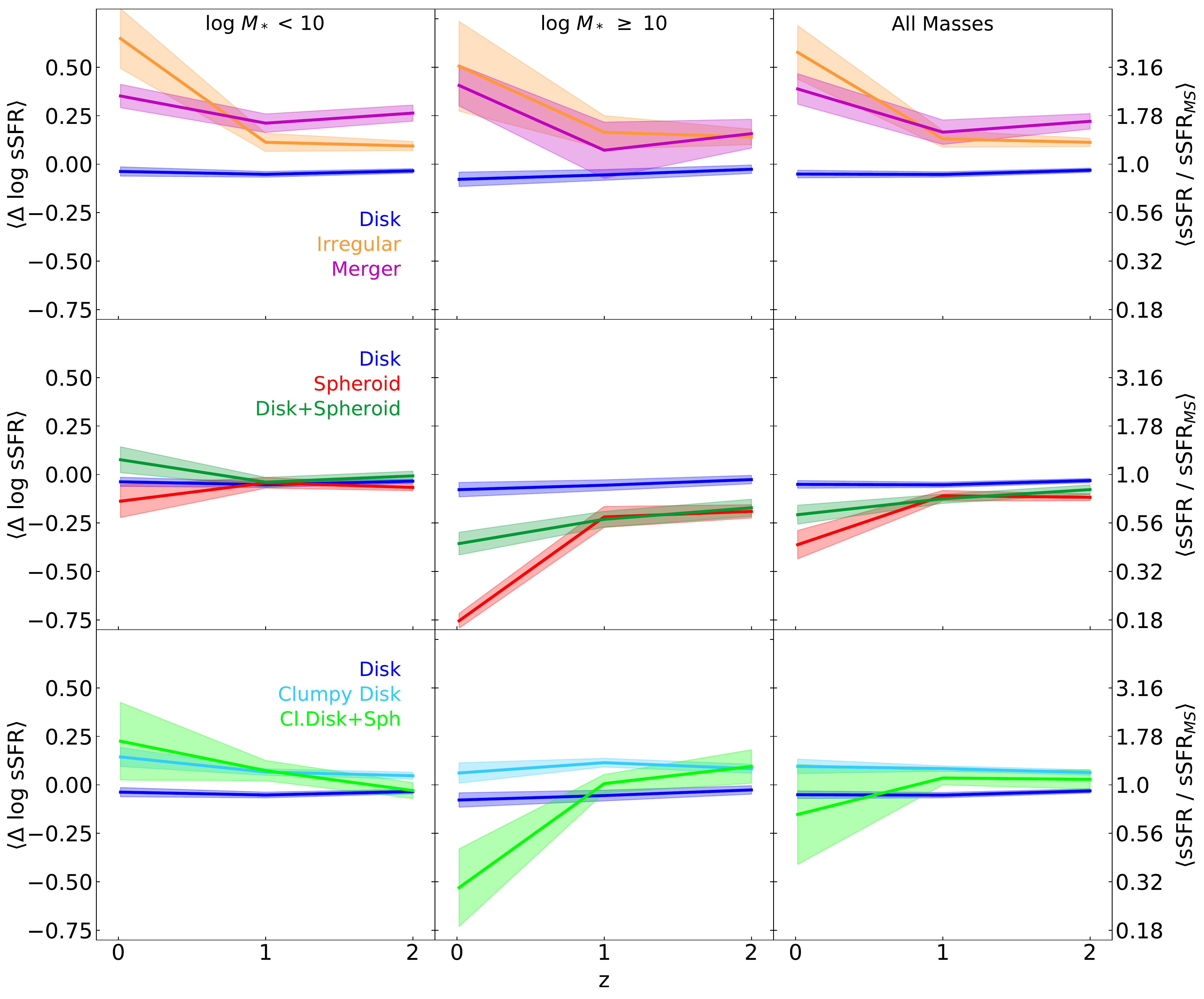}
    \caption{Evolution of the means of the $\Delta$ log sSFR distributions (see Figure \ref{deltaSSFRGauss}) for different morphological classes.  The mean $\Delta$ log sSFR represents the typical enhancement (or deficiency) in SFR with respect to the MS.  Mergers have a net SFR enhancement at all masses and redshifts which may be highest at $z \sim 0$.  Irregulars are enhanced at all redshifts.  Spheroids and Disk + Spheroids are mostly centered on the MS at low mass, but fall well below the MS at recent times at high mass.  Errors were derived by resampling with substitution.  When only a single galaxy is present (e.g., low-mass Clumpy Disk + Spheroids at $z \sim 0$), we instead adopt an error of 0.2 dex, corresponding to a pessimistic estimate of the error in SFR.}
    \label{deltaSSFRMeansvZ}
\end{figure*}

In this section, we take a closer look at the relative sSFR of different morphological classes, focusing on the means of the $\Delta$ log sSFR distributions shown in Figure \ref{deltaSSFRGauss}.  The means represent the typical enhancement in SFR relative to the MS, and are shown versus redshift in Figure \ref{deltaSSFRMeansvZ}.  At low mass, Disks, Spheroids, and Disk + Spheroids show little to no difference relative to the MS at any redshift; this is expected for Disks since they essentially define the MS.  Clumpy Disks have a modest enhancement ($\sim 0.1$ dex) at all redshifts.  Mergers show an SFR enhancement that hovers at $\sim 0.3$ dex.  Irregulars possess a slight enhancement at $z \sim 1$ and $z \sim 2$ which dramatically increases to $\sim 0.7$ dex at $z \sim 0$.

Moving to high mass, we see little change in the behavior of the means (with respect to low mass) for either Disks or Clumpy Disks.  In contrast, Spheroids and Disk + Spheroids have an SFR deficiency of $\sim -0.2$ dex at $z \sim 2$ which decreases significantly with time; Spheroids drop to $\sim -0.75$ dex and Disk + Spheroids drop to $\sim -0.3$ dex by $z \sim 0$.  A similar decrease is seen for Clumpy Disk + Spheroids, but may not be meaningful as the sample size at $z \sim 0$ consists of a single galaxy (see Table \ref{SampleSizeTable}).  This indicates that high-mass spheroidal (disky or pure) galaxies become more strongly associated with the RS over time.  Mergers show the same behavior as at low mass, though the enhancement factors are somewhat lower at $z \gtrsim 1$ than for low masses.  It is important to note that the low number of Mergers included in our sample at $z \lesssim 1$ and the low number of Irregulars at $z \sim 0$ (see Table \ref{SampleSizeTable}) mean that the associated distributions are poorly sampled; this should be kept in mind as a caveat to our $z \lesssim 1$ Merger results.

\subsection{Contribution to the SFR Budget of Different Morphological Classes} \label{Section:Results:MSBudgetResults}

\begin{figure*}[ht!]
    \centering
    \includegraphics[scale=0.44]{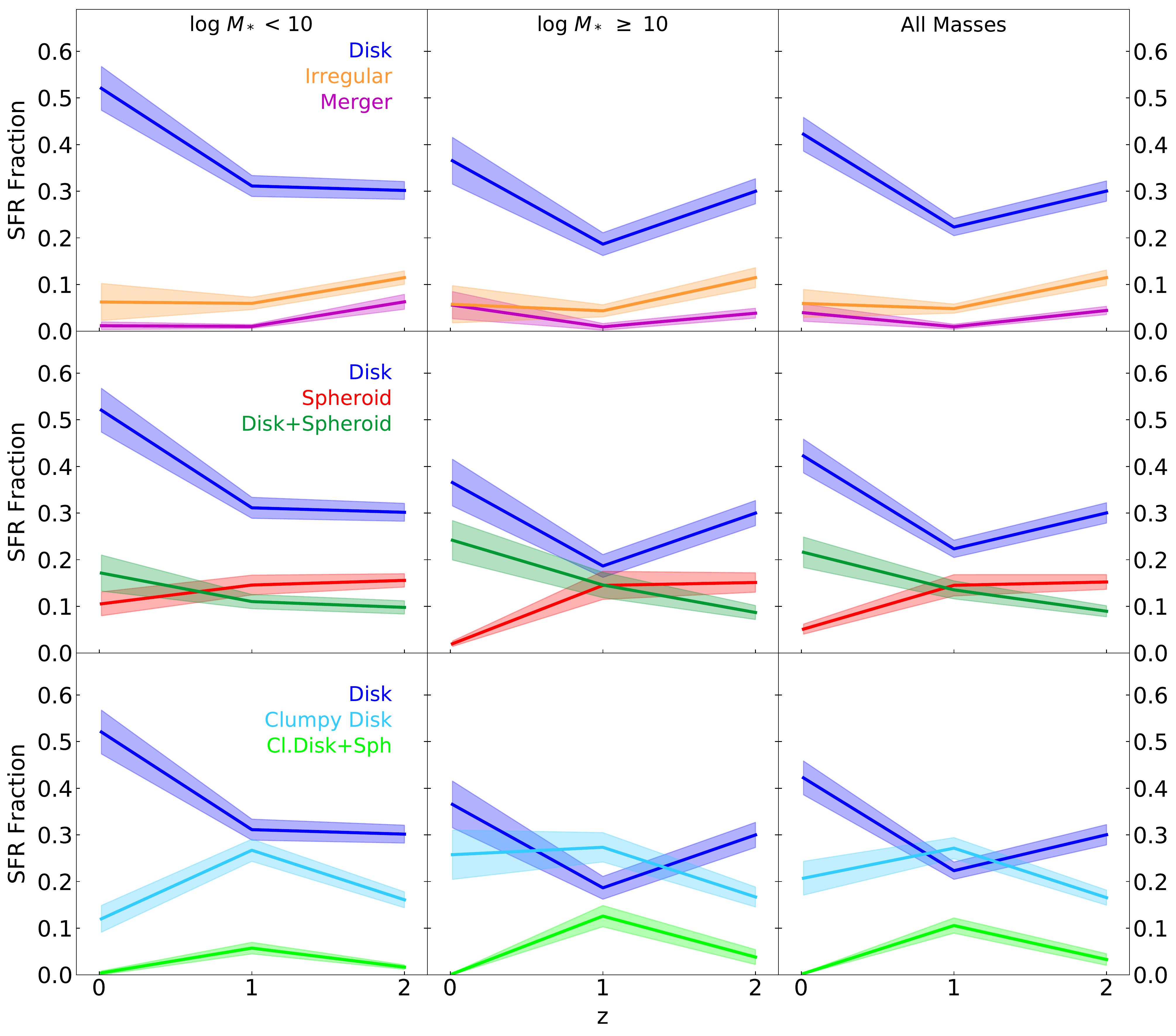}
    \caption{Evolution of the SFR budget for star-forming galaxies.  Disks and Clumpy Disks are the greatest contributors to the SFR budget at all redshifts.  Clumpy Disks and Clumpy Disk + Spheroids both peak at $z \sim 1$ and have greater contributions at high mass.  The total SFR fraction of the clumpy classes is similar at $z \sim 2$ and $z \sim 0$ in each mass bin.  Spheroids show a decline in SFR fraction with time which is most dramatic from $z \sim 1$ to $z \sim 0$ at high mass.  Disk + Spheroids increase in contribution over time, while Clumpy Disk + Spheroids peak at $z \sim 1$ and nearly vanish by $z \sim 0$.  Irregulars show no change in their contribution from $z \sim 1$ to $z \sim 0$ despite dropping in number fraction.  The Merger contribution is minimal at all redshifts, never making up more than $\sim 5\%$ of the budget.  Errors were derived by resampling with substitution.}
    \label{SFRBudgetEvol}
\end{figure*}

In this section, we present the SFR budgets for low- and high-mass star-forming galaxies (SFGs; $\Delta$ log sSFR $\geq -1.0$) with different morphologies.  We show the redshift evolution of these fractions in Figure \ref{SFRBudgetEvol}.  The SFR fractions of Figure \ref{SFRBudgetEvol} are defined as the sum of SFRs for all galaxies of a given class in a specific bin divided by the sum of SFRs for all galaxies within that bin.  At low mass, we see that Disks contribute significantly at all redshifts, becoming the majority contributor ($\sim 60\%$) to the SFR at $z \sim 0$.  Clumpy Disk and Clumpy Disk + Spheroid fractions collectively peak at $z \sim 1$ where they form $\sim 33\%$ of the budget, but drop by $z \sim 0$ where their contribution is $\sim 12\%$.  Spheroids have a sizable contribution ($\sim 15\%$) at $z \gtrsim 1$ which decreases by $z \sim 0$ (to $\sim 10\%$).  Irregulars peak in contribution at $z \sim 2$ ($\sim 10\%$) but have similar contributions ($\sim 5\%$) at both $z \sim 1$ and $z \sim 0$, despite a decrease in their number fraction.  Mergers contribute no more than $\sim 5\%$ of the budget at any redshift.  

The trends at high mass are largely similar to those at low mass, though there are some differences.  The most striking difference is in the contribution of clumpy galaxies at $z \sim 1$, where Clumpy Disks exceed Disks.  Clumpy Disk + Spheroids also show a greater contribution ($\sim 12\%$) at $z \sim 1$.  As at low mass, the collective share of Clumpy Disks and Clumpy Disk + Spheroids peaks at $z \sim 1$ ($\approx 38\%$) but decreases by $z \sim 0$ ($\sim 26\%$).  Interestingly, the share in the budget for Clumpy Disks stays relatively constant from $z \sim 1$ to $z \sim 0$ despite the decrease in their number fractions.  Spheroids show a sharp decline in contribution from $z \sim 1$ to $z \sim 0$, almost disappearing from the budget at low redshift despite only slightly decreasing in number fraction; this is because their typical SFR decreases significantly by $z \sim 0$ (see Figures \ref{deltaSSFRGauss} and \ref{deltaSSFRMeansvZ}).  While the high-mass Merger fractions at $z \sim 0$ and $z \sim 2$ are similar, there is a sharp dip in their SFR contribution at $z \sim 1$; this is because their number fraction (see Table \ref{SampleSizeTable}) as well as their mean $\Delta$ log sSFR (see Figure \ref{deltaSSFRMeansvZ}) are lowest at $z \sim 1$.  

\subsection{Morphological Composition of the Red Sequence}
\label{Section:Results:RedSequenceComposition}

\begin{figure}
    \centering
    \includegraphics[scale=0.4]{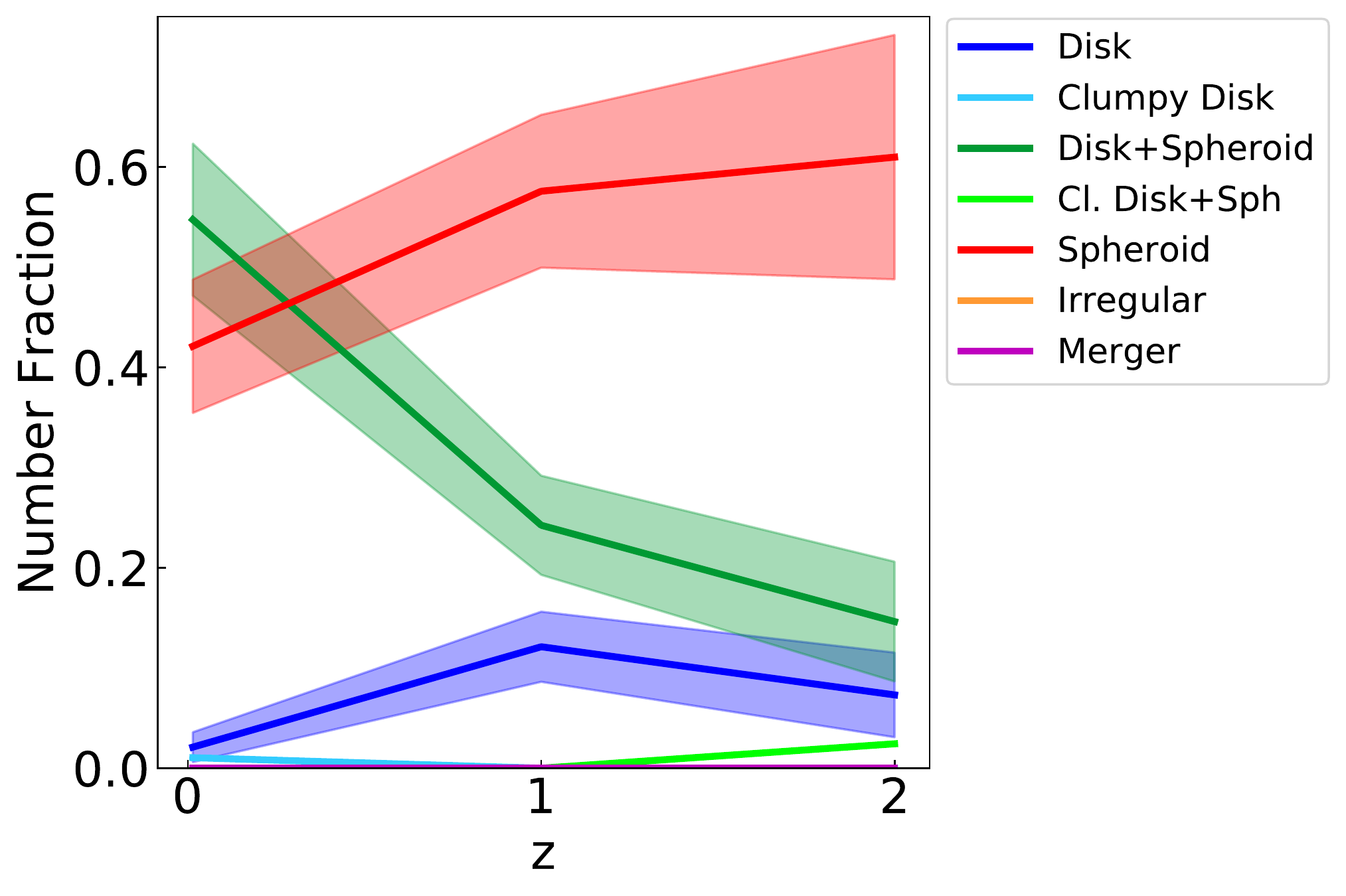}
    \caption{Evolution of the red sequence (RS) number fractions.  The error regions are determined by the Poisson error ($\sqrt{N}$) of the number counts, and are shown only for Disks, Spheroids, and Disk + Spheroids.  Spheroids and Disk + Spheroids dominate at all redshifts, though normal Disks are found at $z\sim1$ and $z \sim 2$ as well.  Although the RS at $z \sim 2$ is comprised largely of Spheroids, by $z \sim 0$ the Disk + Spheroids have become the majority.  Contributions from other classes are negligible.}
    \label{RSFracEvol}
\end{figure}

In this section, we look at the evolution of the morphological composition of red sequence (RS) galaxies.  Our RS samples consist of all galaxies with $\Delta$ log sSFR $< -1.0$; the RS label is colloquial as we do not perform any color-based selections.  Owing to the less accurate SFRs of RS galaxies, we use only the number of galaxies in a class to determine its relative contribution to the RS.  We show the redshift evolution of the number fractions for galaxies on the RS in Figure \ref{RSFracEvol}.  We find that the RS is primarily composed of Spheroids and Disk + Spheroids at all redshifts.  Disks comprise roughly $10\%$ of the RS on average while the other classes contribute negligibly, if at all.  Notably, we see that the fraction of Spheroids drops with redshift while the fraction of Disk + Spheroids increases; this is more dramatic for Disk + Spheroids, which increase from about $25\%$ at $z \sim 1$ to around $55\%$ at $z \sim 0$.  

We also check the number fractions for green valley (GV) galaxies ($-1.0 < \Delta$ log sSFR $< -0.45$) and find the same trends; lower Spheroid fractions and higher Disk + Spheroid fractions over time.  The Disk and Disk + Spheroid fractions are also much higher on the GV than on the RS at all redshifts, indicating that the GV is `diskier' than the RS throughout cosmic time; this is consistent with other literature results \citep{Mendez2011AEGISGreenValley, Lee2018GalaxyStructure}.  The fraction of GV galaxies (relative to the total sample at each redshift) is $\sim 16\%$ at $z \sim 0$, $\sim 6\%$ at $z \sim 1$, and $\sim 5\%$ at $z \sim 2$.

\section{Discussion} \label{Section:Discussion}

We will focus our discussion on three classes of particular interest in the study of galaxy evolution:  mergers, clumpy disks, and spheroids.  

\subsection{Mergers}

The galaxies that we and \citetalias{K15} identify as mergers tend to be in the latest interaction stage, where the interacting galaxies have coalesced.  This is similar to the `post-mergers' of \citet{Ellison2013PostMergers}, who tracked the SFR enhancement of local ($0.005 < z < 0.1$) interacting galaxies across different stages in the merging process.  They found that, on average, the highest induced SFR occurs in the post-mergers, with the largest SFR enhancement located in the galaxy center.  We show in Figure \ref{deltaSSFRMeansvZ} the Merger SFR enhancements with respect to the MS line, taken to be the mean of the $\Delta$ log sSFR distribution.  Errors are derived via 1,000 bootstrap resamplings of the Merger $\Delta$ log sSFR distributions for each redshift and mass bin.  Our enhancement factors are broadly consistent with \citet{Ellison2013PostMergers} (see their Figure 6) as well as other studies investigating the SFR enhancement in merging galaxies at different redshifts \citep{Bridge2010Mergers, Kaviraj2015MergerSim, Knapen2015Interactions, Martin2017MergerCont, Fensch2017HighzMergerSim}.  

Studies of simulated mergers have found that the SFR enhancements in merging pairs decreases with increasing gas fraction, suggesting that mergers of high-redshift galaxies, which are more gas-rich, should have lower SFR enhancements than their low-redshift counterparts.  This has been confirmed for galaxy pairs at $z \sim 2$, but not for late-stage or post-merging galaxies \citep{Wilson2019SFRandZinMergersatHighz}.  The simulated post-mergers of \citet{Hani2020TNGPostMergerSims} show a constant SFR enhancement (a factor of $\sim 2$, or $\sim 0.3$ dex above MS) across the redshift range $0 < z < 1$, which may indicate that the late-stage SFR enhancements remain high out to $z \sim 2$.  Our Mergers are enhanced at all redshifts, with the highest enhancement at low redshift ($z \sim 0$, see again Figure \ref{deltaSSFRMeansvZ}), though the differences are not extreme for low or all masses where the errors are lowest.  The Merger enhancement factors at $z \gtrsim 1$ are even lower when using the SFRs and masses of \citet{Fang2018UVJCANDELS}, providing stronger support for a decreasing enhancement with increasing redshift.  It should be noted, however, that for $z \lesssim 1$ our Merger sample size is very low (only a few galaxies in each bin, see Table \ref{SampleSizeTable}); thus our results for $z \lesssim 1$ Mergers are not definitive.

We find that Mergers contribute $\lesssim 5\%$ to the SFR budget for all masses and redshifts (see Figure \ref{SFRBudgetEvol}).  Similar results have been obtained for major mergers (typically with a mass ratio $<$ 4:1) using a variety of selection methods for a wide range of redshifts \citep[][]{Robaina2009MergerSFRContribution, DePropris2014Mergers, Lofthouse2017MajorMergers}, as well as simulations \citep{Kaviraj2015MergerSim, Rodriguez2019SimbaSim}.  Inferring the mass ratio of merger progenitors from the visible tidal features is very challenging and has not been explored extensively in the literature so far.  We therefore cannot divide our sample into major and minor mergers.  Even so, the consistency of our results with major merger studies and the expectation that galaxies of similar mass should produce more visible merger signatures leads us to assume that the selection used in this work, obtained from or following the \citetalias{K15} classifications, captures major mergers.  

The fraction of star formation attributable to major mergers will depend on which of the phases of the merging process are classified as mergers.  Indeed, \citet{Puech2014ReinstatingMajorMergers} point out that if one attributes pre-fusion, fusion and post-merger phases under the merging label, the total SF attributable to these phases can exceed $50\%$ at $z \sim 0.6$.  However, most of this star formation would have occurred in these galaxies even without any interaction, so a more limited label, such as what is used in this study may better convey the contribution of the merger process to the SFR budget.  Notwithstanding these differences in labelling, our lower estimate is nevertheless consistently low at different redshifts, suggesting that whatever the total contribution of mergers may be, it has not been significantly higher at previous cosmic epochs.  

Minor mergers (mass ratios $>$ 4:1) may play a more significant role, as evidenced by \citet{Kaviraj2014MinorMergersStripe82} who use an indirect method to estimate that 1/4 of the star formation in $z < 0.07$ late-type galaxies is attributable to minor mergers (mass ratios $>$ 4:1). However, their estimate is based on assuming that the star formation observed in some ETGs is entirely due to minor, gas-rich mergers, in disagreement with the studies that find that the majority of ETGs with star formation owe it to low-level continuous star formation \citep{Fang2012ExtendedSFinMidzETGs, Salim2012SFinOpticallyRedETGs}.

It should be noted that our results cannot be used to directly constrain merger rates.  Recent works have demonstrated that the merger observability timescale is likely to be shorter at earlier times, leading to underestimation of the merger rate at higher redshifts when using visual classifications \citep[see][and references therein]{Lotz2011MergerRates, Snyder2017MassiveClosePairsHighz}.  Because of this, and because our Mergers represent a specific stage in the merging process, our Merger fractions are only loosely representative of the merger rates. 

\subsection{Clumpy Disks}

One of our more curious results is the apparent increase in clumpy galaxy fractions from $z \sim 2$ to $z \sim 1$, since most other studies find that the fraction of clumpy galaxies is highest at $z \sim 2$ \citep{Guo2015ClumpsI, Shibuya2016ClumpyGalaxies}.  We also find a greater incidence of clumpy galaxies at high mass than at low mass at all redshifts.  This is seen in Figures \ref{deltaSSFRGauss}, \ref{TypeFracEvol}, and \ref{SFRBudgetEvol} where Clumpy Disks and Clumpy Disk + Spheroids have greater combined contributions at high masses.  Previous studies have found an increase in the size of disks from high to low redshifts and from low to high masses \citep{Wuyts2011StructureEvolution, vanderWel2014GalaxyMassSizeEvol, Margalef-Bentabol2016}.  Thus, a possible explanation for our results is that it is easier to `fit' a clump inside a larger, more massive disk.  This may also explain why our Clumpy Disk fractions at $z \sim 0$ are similar to those at $z \sim 2$.  Alternatively, larger clumps may be preferentially identified by human classifiers, leading to underestimates of the clumpy fraction among small or low-mass galaxies with correspondingly smaller or lower-mass clumps.  

The increase in clumpiness we find from $z \sim 2$ to $z \sim 1$ may be related to the different processes from which clumps are formed.  Clumps may be formed via violent disk instabilities (VDI) induced by the accretion of cold cosmic gas \citep[i.e.,][]{Dekel2009Accretion&VDI, Ceverino2010ClumpySims, Cacciato2012Accretion&VDI, Oklopcic2017GiantClumpsinFIRE;ClumpyCaseStudy} or via mergers \citep{Puech2010ClumpyGalaxies, Guo2015ClumpsI, Zanella2019Clumps}.  Minor mergers are an appealing explanation for the peak in clumpy galaxy fractions we observe at $z \sim 1$, as cold gas accretion should be less significant than at $z \sim 2$ \citep{Dekel2009Accretion&VDI, Oklopcic2017GiantClumpsinFIRE;ClumpyCaseStudy}.  Major mergers may also contribute to clump formation \citep{Ribeiro2017VIMOSTwoClumpSystemsFromMajorMergers}, but they are less frequent \citep[see][]{Lotz2011MergerRates, Mantha2018MergerRate, Duncan2019MergerRates} and may simply destroy the disk, though this is not guaranteed \citep{Lotz2008MergerMorphology&TimescalesSims, Sparre2017MergersIntoSpirals, Martin2018MorphTransformSims}.  We do not necessarily expect the minor merger rate to rise from $z \sim 2$ to $z \sim 1$, as our results might suggest, though minor merger rates are highly uncertain even at low redshift \citep[see][]{Lotz2011MergerRates, Duncan2019MergerRates}.  The minor merger explanation is called into question by recent results, however, which find that clump masses for CANDELS star-forming galaxies are largely consistent with an in-situ formation scenario (Huertas-Company et al. 2020, in prep).  This may suggest that our results arise due to the aforementioned size effect, where larger galaxies have correspondingly larger and brighter clumps which are preferentially identified by human classifiers. 

\citet{Guo2015ClumpsI} used an automated `blob finder' to identify star-forming regions in the \textit{HST}/ACS images for 3,239 log $M_* / M_{\odot}$ $<$ 10.6 galaxies in CANDELS (GOODS-S and UDS fields) at $0.5 < z < 3$.  They defined clumps as blobs which contribute more than $8\%$ of the total UV light of their host galaxies.  In contrast to our results, they find that a much higher fraction of SFGs are clumpy (as much as $\sim 60\%$ at $z \sim 2$), and also that higher mass bins have lower clumpy fractions.  It is worth pointing out that clumpy galaxy fractions are highly sensitive to methodology and clump definition and vary widely in the literature \citep[e.g.,][]{Ravindranath2006ClumpyLBGs, Elmegreen2007HUDFClumpyDisksExtremeRedshift, Wuyts2012ClumpyGalaxies, Guo2015ClumpsI}, so we should not expect complete agreement.  Comparing their observed clumpy galaxy fractions (as a function of redshift) to fractions derived from the \citetalias{K15} classification scheme, \citet{Guo2015ClumpsI} find that their results agree best with clumpy fractions derived using both the clumpiness and patchiness flags from the \citetalias{K15} data release (see their Appendix A) rather than either the clumpy or patchy flags alone.  This is because the blob-finder does not account for the light concentration of the blobs.  The inclusion of patches may help explain why their clumpy fractions are generally higher than ours.  \citet{Guo2015ClumpsI} also exclude very small ($< 0.1\arcsec$) and elongated (axis ratio $< 0.5$) galaxies from their sample.  Our inclusion of such galaxies could easily lead to lower clumpy fractions given that many galaxies with half-light radius $< 0.1\arcsec$ are Spheroids (see Figure \ref{BiasCheck}), which rarely possess clumps.  We also include edge-on disks whose clumps may be obscured by dust.  The inclusion of galaxies with unresolved or obscured clumps may imply that we are underestimating the clumpy fractions.  However, we do include the contribution from non-disky compact or irregular SFGs which would be excluded by the \citealt{Guo2015ClumpsI} cuts.  The contribution from such galaxies is not insignificant, especially at $z \sim 2$, so our looser selection is not without merit.  Even if we do underestimate our clumpy fractions, our consistent sample selection and methodology ensures that they should be similarly underestimated at all redshifts, preserving the general evolutionary trends. 

In any case, our results suggest that disk evolution is an ongoing process at $z \sim 1$.  This is supported by kinematic studies, which find a gradual decrease in disordered gas kinematics in disk galaxies from $z \sim 2$ to $z \sim 0$ \citep[e.g.,][]{Kassin2012DiskSettling}.  The downsizing phenomenon noted by both \citet{Kassin2012DiskSettling} and \citet{Guo2015ClumpsI}, wherein lower-mass galaxies are more disordered in their gas motions and more clumpy in appearance, is conspicuously absent in our results.  As our clumps are visually identified, our sample may be biased towards larger and more massive or more concentrated clumps.  

We find substantially higher clumpy galaxy fractions among local SFGs than the current literature suggests.  Previous studies of clumpy galaxies at low redshifts ($z < 0.5$) have typically been limited to local analogs of high-redshift galaxies \citep[][]{Overzier2009ClumpyLowzLBAs, Elmegreen2013ClumpyGalaxiesinKiso&UDF, Fisher2017DYNAMOClumpsLocalAnalogs}.  The first systematic search for low-redshift clumpy galaxies was performed by \citet{Murata2014ClumpsinCOSMOS}, who used an automated algorithm to identify clumpy galaxies in the redshift range $0.2 < z < 1$ using \textit{HST}/ACS F814W imaging of the COSMOS field.  They found that the fraction of SFGs with clumps at $0.8 < z < 1$ is $\sim 35\%$, in good agreement with our results (see Figure \ref{TypeFracEvol}).  At lower redshifts, however, the F814W filter shifts into the rest-frame optical where clumps are less visible.  Indeed, at $0.2 < z < 0.4$, they find a fraction of only $\sim 5\%$, lower than our $z \sim 0$ fractions ($\sim 10\%$ at low mass and $\sim 15\%$ at high mass).  As previously discussed, our sample selection and visual classification may underestimate the fraction of clumpy galaxies, especially among smaller or lower-mass galaxies.  In that case, the number and SFR contribution of Clumpy Disks at $z \sim 0$ may be even higher.  

Using visual classifications for 1213 UDS galaxies at $1 < z < 3$, \citet{Mortlock2013HubbleSequenceFormation} find that the general population of log $M_* / M_{\odot} > 10$ galaxies at $z \gtrsim 1.86$ is largely peculiar (i.e., irregular) or spheroidal with a minimal ($\sim 0\%$) fraction of disks.  This conflicts with Figure \ref{TypeFracEvol}, which shows a substantial ($\sim 50\%$) disk population at $z \sim 2$.  The discrepancy in disk fractions is likely a result of classification scheme differences.  Supporting this is the observation that many of the disks in our $z \sim 2$ sample possess disturbed morphologies which may lead to a peculiar classification under the \citet{Mortlock2013HubbleSequenceFormation} scheme; an example of this is the Clumpy Disk of Figure \ref{ClassImages}.  Indeed, \citet{Mortlock2013HubbleSequenceFormation} interpret their results as an absence of the traditional `settled' disks observed at low redshifts rather than an absence of any disks at $z \gtrsim 1.86$.  Furthermore, \citet{Mortlock2013HubbleSequenceFormation} classify all interacting galaxies as peculiar, while we only consider interaction status when assessing whether a galaxy is a Merger, leading to more galaxies classified as disks under our scheme.  Despite the discrepancy in disk fractions, we nonetheless find some agreement with \citet{Mortlock2013HubbleSequenceFormation} in that the number fraction of Irregulars increases steadily with increasing redshift (see again Figure \ref{TypeFracEvol}).

\subsection{Spheroids} \label{Discussion:Spheroids}

We observe significant evolution in the relative contribution of Spheroids to the SFR budget of both star-forming galaxies (SFGs) and the `red sequence' (RS).  SFG Spheroids are common at all redshifts, especially at $z \gtrsim 1$, suggesting that the process driving morphological transformation (i.e., from a disk to a spheroid) precedes the process that drives quenching of star formation.  Evidence for the distinction between these two processes has been found in other studies as well \citep[see][]{vanderWel2011CmpctMassiveGalsAreDiskDomAtCosmicNoon, Barro2013compactSFGsCANDELS, Tacchella2015InsideOutQuenching, Brennan2015Quenching&MorphTransform, Tacchella2018DustBulges&Quenching, Koyama2019Morphology&SFE}.  Some galaxies may skip the disk phase entirely, however, and assemble as star-forming Spheroids directly \citep{Cappellari2016FastVsSlowRotatingEllipticals, Tacchella2019IllustrisTNGDiskSpheroidBuildup}.  

Spheroidal SFGs, sometimes referred to as `blue nuggets', are expected to form via episodes of gas compaction induced by either secular processes or mergers \citep{Tacchella2016Compaction&ConfinementInTheSFMS, Zhang2019ProlateToDiskyEvol}.  Blue nuggets were initially predicted by simulations and then confirmed to exist by observational studies \citep[][]{Lang2014BulgeGrowth&Quenching, Nelson2014MassiveCompactFormingCoreGalaxy, Barro2014CompactHighzProgenitorsOfCompactQuiescGals, Zolotov2015Compaction, Tacchella2016HighzCompaction&Quenching, HuertasCompany2018BlueNuggetsML}.  Following this blue nugget phase, feedback from star formation and/or AGN activity, as well as buildup of the halo, may then lead to quenching \citep[see][]{Brennan2015Quenching&MorphTransform, Tacchella2016HighzCompaction&Quenching}.    

The blue nugget compaction scenario is further supported by \citet{Zhang2019ProlateToDiskyEvol}, who find that galaxies evolve in their intrinsic shapes from prolate to oblate over time.  Their results suggest that this transformation occurs via compaction in a characteristic mass range which decreases with redshift, from log$(M_*/M_{\odot}) \sim 10$ at $z \sim 2$ to log$(M_*/M_{\odot}) \sim 9.3$ at $z \sim 0.75$.  It may be that our star-forming Spheroids correspond to the blue nuggets expected to be produced by the prolate-oblate transformation.  Indeed, the evolution in mean $\Delta$ log sSFR shown in Figure \ref{deltaSSFRMeansvZ} appears to mirror the results of \citet{Zhang2019ProlateToDiskyEvol}.  At $z \sim 2$, where Spheroids have a mean $\Delta$ log sSFR close to the MS, the transformation mass corresponds roughly to our separation between high and low masses.  The high-mass Spheroid distribution mean then decreases significantly until $z \sim 0$ (see also Figure \ref{deltaSSFRGauss}), paralleling the decrease in transformation mass.  However, although the transformation mass should be well below our lower mass limit of log$(M_*/M_{\odot}) = 9.3$ by $z \sim 0$, the low-mass mean for Spheroids remains mostly unchanged over time.  Despite this, the low-mass Spheroid distribution still evolves over time, becoming broader with a much weaker peak and developing a tail towards the RS by $z \sim 0$ (see again Figure \ref{deltaSSFRGauss}).  Our results are therefore largely consistent with the evolution in intrinsic galaxy shapes over time suggested by \citet{Zhang2019ProlateToDiskyEvol}. 

The number and SFR contribution of SFG Spheroids are nearly constant from $z \sim 2$ to $z \sim 1$ at both high and low mass, but by $z \sim 0$ SFG Spheroids decrease significantly in SFR contribution at high mass (see Figure \ref{SFRBudgetEvol}).  The fraction of Spheroids on the RS also decreases over time, with the RS comprising mostly of Disk + Spheroids at $z \sim 0$ (see Figure \ref{RSFracEvol}).  This may be due to the formation of disks around Spheroids, but could also be due to the growth of significant bulges (i.e., spheroids) within disks, corresponding to field S0 galaxies.  The shift in RS fractions may imply a transition in the dominant mode of quenching, where at $z \sim 2$ the process is relatively quick while at $z \sim 0$ the quenching is more passive and less likely to destroy a disk.  Evidence for such a shift in the quenching mechanism has been found by other studies \citep[e.g.,][]{Cassata2013ETGs, HuertasCompany2016, Chen2019BH&HaloQuenching}.

%Violent disk instabilities (VDIs) triggered by mergers, accretion, or secular processes may also explain the existence of starbursting spheroids at high redshift \citep{Zolotov2015Compaction, Tacchella2016HighzCompaction&Quenching}.  

Using bulge-disk decomposition, \citet{Margalef-Bentabol2016} found that at $z > 2$ pure disks and pure bulges (i.e., spheroids) dominate, while at $z < 2$ the number density of two-component galaxies (possessing an outer disk and inner bulge) increases to match that of pure galaxies.  They also find that disks undergo significant size evolution in this period while the bulge size remains mostly constant.  \citet{Sachdeva2019DiskFormation}, also using bulge-disk decomposition, found that the number of pure spheroids drops more rapidly than the number of pure disks from $z > 2$ to $z < 2$.  They conclude that disk growth around pre-existing spheroids is a viable channel for morphological transformation at $z \sim 2$.  It should be noted that a pure spheroid, as determined by the 2-component Sersic fitting of \citet{Sachdeva2019DiskFormation}, includes all galaxies whose light profiles are best fit by a Sersic profile regardless of the Sersic index.  Galaxies with low Sersic indices (i.e., a less concentrated light distribution) would likely be classified as disky under our visual scheme.  Our non-spheroidal disk classes may also contain two-component disk+bulge systems where the bulge is not visually prominent, so a one-to-one comparison between our results and these studies is challenging.  Even so, the general consistency between our results is encouraging.  

\begin{figure}
    \centering
    \includegraphics[scale=0.35]{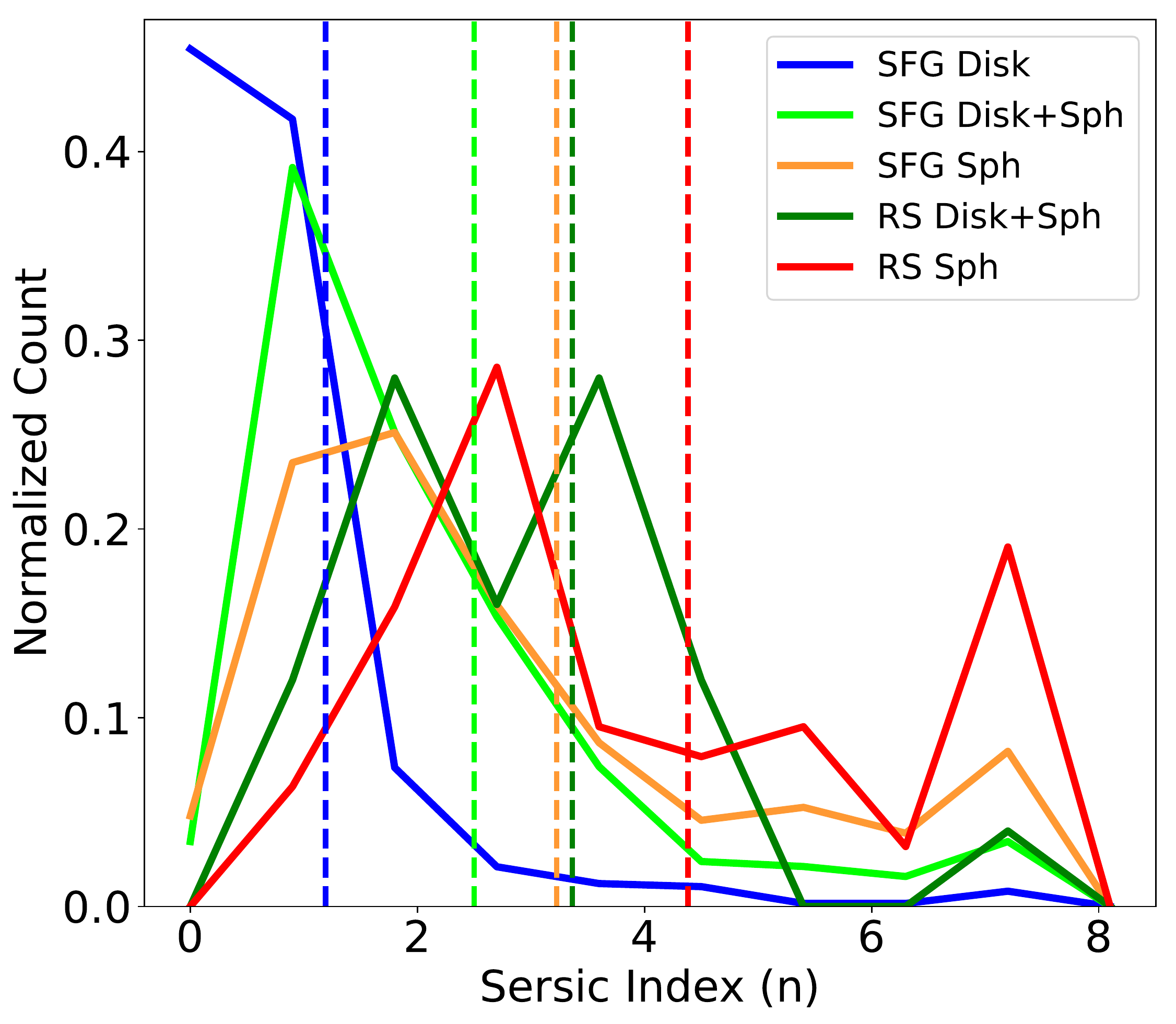}
    \caption{Normalized histograms of the distribution of \textit{H}-band Sersic indices for different classes of star-forming (SFG) and red sequence (RS) galaxies at $z \sim 1$ and $z \sim 2$.  The dashed lines mark the average Sersic index of each class.  RS galaxies have higher Sersic indices on average than their SFG counterparts, regardless of class.  Notably, the average Sersic index for SFG Spheroids is n $\sim 3$ while for RS Spheroids it is n $\sim 4$.  This suggests that SFG Spheroids are less concentrated than RS Spheroids.}
    \label{SersicCompare}
\end{figure}

In order to better understand the characteristics of our Spheroids, we show in Figure \ref{SersicCompare} the \textit{H}-band Sersic index distributions for different spheroidal subsets of our GOODS-S ($z \sim 1$ and $z \sim 2$) galaxies.  Of particular interest are the distributions of star-forming Spheroids (SFG Spheroids) compared to the RS Spheroids.  From Figure \ref{SersicCompare} it can be seen that RS Spheroids have an average index of about 4, while SFG Spheroids have an average index of around 3, comparable to RS Disk + Spheroids but higher than SFG Disk + Spheroids.  Other studies have found that the average Sersic index for star-forming galaxies is lower than the average for quiescent galaxies \citep[e.g.,][]{Wuyts2011StructureEvolution}.  Notably, Figure \ref{SersicCompare} shows that this trend persists even when considering only Spheroids (and Disk + Spheroids as well).  Using imaging data from the HUDF, \citet{Elmegreen2007HUDFClumpyDisksExtremeRedshift} also found a large range in Sersic indices for visually-identified ellipticals across the redshift range $0 \lesssim z \lesssim 5$), with a distribution centered at $n \sim 1.5$.  Our Spheroid class is therefore less homogeneous than our broad classification scheme suggests.  

%Several other studies of high-redshift ellipticals (i.e., spheroids) have also found that a substantial fraction of ellipticals have $n < 4$ (Citation needed).

Using a Convolutional Neural Network (CNN) trained on the \citetalias{K15} visual classifications, \citet{HuertasCompany2016} studied the evolution of different morphological classes with redshift for 50,000 log $M_* / M_{\odot}$ $> 10$ galaxies across all 5 CANDELS fields.  The CNN assigned synthetic vote fractions to each galaxy for the five main morphological classes (see \citetalias{K15}), which they used to classify galaxies in a similar manner to our study, though limited to four classes:  disks, spheroids, disk + spheroids, and irregulars.  Though we find generally good agreement with their results, in their lowest redshift sample ($0.2 < z < 0.5$) they find that at very high masses (log $M_* / M_{\odot}$ $\gtrsim 10.7$) the SFGs are dominated (in number) by Disk + Spheroids while the RS is dominated by Spheroids.  This contrasts with our results; even when only considering galaxies above log $M_* / M_{\odot}$ = 10.7, we find that Disks still dominate the SFGs and Disk + Spheroids still dominate the RS at $z \sim 0$.  Despite this, we nonetheless find that the fraction of Disk + Spheroids (normal and clumpy) increases over time, and is greater at high mass relative to low mass at each redshift (see Figure \ref{TypeFracEvol}).  We also find good agreement with other studies in the local universe.  In particular, using a more inclusive redshift window than ours ($z < 0.075$), \citet{Lofthouse2017LocalSFRBudget} found that the local SFR budget for high-mass (log $M_* / M_{\odot}$ $>$ 10) galaxies is dominated ($\approx 65\%$) by disks with low bulge-to-total ratios ($<$ 0.25), consistent with our results. 

\section{Conclusions} \label{Section:Conclusions}

In this paper, we combine imaging from \textit{HST}, SDSS, and \textit{GALEX} to form a sample of $\sim$ 3,000 galaxies at $z \sim 0$ ($z \sim 0.01$), $z \sim 1$ ($0.8 < z < 1.4$), and $z \sim 2$ ($1.5 < z < 2.5$).  Using the \citet{K15} (K15) catalog of visual classifications, we re-classify $z \sim 1$ and $z \sim 2$ galaxies into mutually exclusive classes which include Mergers and Clumpy Disks alongside more traditional classes (Disk, Spheroid, Irregular).  With our \citetalias{K15}-matched sample as a training set, we use \textit{GALEX} UV and SDSS optical images (degraded to match the physical scale of \textit{HST} images) to classify our $z \sim 0$ sample in a manner consistent with \citetalias{K15}.  A consistent SED fitting method is used to derive SFRs which includes a flexible dust attenuation law.  We quantify the distribution of different classes of star-forming galaxies (SFGs) for both low- (log $M_* / M_{\odot}$ $<$ 10) and high-mass (log $M_* / M_{\odot}$ $\geq$ 10) galaxies separately.  For both mass bins we also track the evolution in the SFR budget contributions for different classes of SFGs.  The relative number fractions for different classes are used to quantify the evolution of the low-sSFR red sequence (RS) with redshift.  Our main conclusions are as follows: 

\begin{enumerate}

    \item The morphological composition of star-forming galaxies (SFGs) at $z \sim 1$ and $z \sim 2$ is more diverse than at $z \sim 0$.  At $z \sim 0$ the SFGs are composed almost entirely of disky classes, while at higher redshifts Spheroids, Irregulars, and Mergers are more represented.

    \item Clumpy Disks and Clumpy Disk + Spheroids peak in number and SFR contribution at $z \sim 1$, not at $z \sim 2$ where the rate of cold cosmic gas accretion and (possibly) merger activity peak.  The more recent peak may be the result of size evolution, with bigger galaxies possessing larger and/or more massive (thus more visually prominent) clumps, but in any case suggests that disks continue their principal assembly epoch at $z \sim 1$.  Clumpy Disks remain a substantial contributor even at $z \sim 0$, in both number and star formation, especially at high masses where they contribute $\sim 25\%$ to the SFR budget.  We also find higher fractions of clumpy galaxies at high mass compared to low mass at all redshifts.  
    
    \item Star-forming Spheroids are common ($\gtrsim 10\%$) at all redshifts, especially for $z \gtrsim 1$, suggesting that morphological transformation precedes quenching, as highlighted in previous studies.  On the RS, the Spheroid fraction drops with decreasing redshift, with Disk + Spheroids dominating the RS by $z \sim 0$.  The shift in RS fractions may also be indicative of a transition in primary quenching mechanism, with a more secular process dominating at later times.  Spheroids display a wide range of Sersic indices ($n$).  Star-forming Spheroids have an average Sersic index ($n \sim 3$) which is lower than that of their RS counterparts ($n \sim 4$).  
    
    \item Mergers contribute little to the star formation budget ($\lesssim 5\%$) at all redshifts.  Mergers remain uncommon even among the starbursts ($\Delta$ log sSFR $\gtrsim 0.5$), never making up more than $\sim 8\%$ of such galaxies at any redshift.  Our classification is more sensitive to late-stage mergers with clear interaction signatures, likely major mergers.  Mergers are enhanced in SFR relative to the MS by an average factor of two.  Irregulars have similar enhancements in SFR and contributions to the SFR budget as the Mergers.  
    
\end{enumerate}

The current study is limited by the relatively small number of galaxies, especially in some morphological classes.  Application of our methodology to a larger sample size has the potential to alleviate these limitations. 

\appendix

\section{Optical Image Degradation for Low-Redshift Galaxies}
\label{Appendix:ImageDeg}

In this section we describe, in detail, the process and physical reasoning used in the degradation of SDSS optical images used in our study.  The resolution scale at $z = 2$ ($z = 1$) is 8.596 (8.156) kpc/$\arcsec$\footnote{Ned Wright's Cosmology Calculator \citep{NedWright2006CosmoCalc} was used for this calculation.}.  The PSF FWHM for the \textit{HST} \textit{V}-band and \textit{H}-band is $0.08\arcsec$ and $0.18\arcsec$, respectively.  The physical resolution of a $z = 2$ galaxy is therefore $\approx 0.7$ kpc in the \textit{V}-band and $\approx 1.5$ kpc in the \textit{H}-band.

Our low-redshift images have $4.0 \arcsec$ resolution in FUV (GALEX) and $1.4 \arcsec$ resolution in the optical (SDSS).  To get a comparable physical resolution to \textit{HST} in the rest-frame UV we would need to use FUV images at $z = 0.0085$, which is slightly lower than the GSWLC limit of $z = 0.01$.  We therefore use galaxies in the range $0.01 < z < 0.0176$, with the upper limit chosen to ensure that the sample size is comparable to that at higher redshifts.  At redshift $z \sim 0.01$, however, the SDSS images have a physical resolution that is $\sim 5$ times better than 1.5 kpc for HST.  We therefore degrade the SDSS images to a target scale of 1.5 kpc using Astropy's \citep{Astropy2013, Astropy2018} Gaussian convolution and block reduction to smooth and resample the images, respectively.  First, we smooth the images using a 2D Gaussian kernel of $\sigma = 5.4$ pixels (2.16$\arcsec$, in SDSS images).  We verify that the target resolution is achieved by fitting radial profiles to point sources in the degraded images and measuring their FWHM.  The images are then resampled such that the physical pixel scale (in kpc/pixel) matches that of the \textit{H}-band images.  The pixel scale for the \textit{H}-band (\textit{V}-band) images is 0.06 (0.03)$\arcsec$/pixel, corresponding to $\approx 0.5$ ($\approx 0.25$) kpc/pixel at $z \sim 2$.  The pixel scale of the SDSS optical images is 0.4$\arcsec$/pixel, or $\approx 0.11$ kpc/pixel at $z \approx 0.014$.  To achieve the same physical pixel scale therefore requires a factor of 5 reduction in the pixel count.  Figure \ref{LocalConvExample} shows the effects of this degradation for a galaxy in our sample.  

In the initial resampling of the SDSS images we mistakenly assumed an \textit{H}-band pixel scale of 0.03 $\arcsec$/pixel and applied only a factor of 2 in pixel reduction to the images before using them for classification.  However, as we smooth our images to a PSF FWHM of 13.5 pixels, the level of detail is essentially the same regardless of whether the reduction factor is 2 or 5.  We verified this with visual inspection and conclude that the resampling error has no impact on the classifications.  

\section{The K15 Raw and Fractional Catalogs}
\label{Appendix:K15Catalogs}

\begin{deluxetable}{ccccccc}[ht]
\tablecaption{Example vote fractions for galaxies assigned to different morphology classes.  The clumpy vote fraction is derived from the \citetalias{K15} raw catalog while the others are from the fractional catalog.}
\tablecolumns{7}
\tablenum{2}
\label{ClassTable}
\tablewidth{0pt}
\tablehead{
\colhead{K15 ID} &
\colhead{Our Class} &
\colhead{Spheroid} &
\colhead{Disk} & \colhead{Irregular} & \colhead{Clumpy} & \colhead{Merger}
}
\startdata
ers2\textunderscore12344 & Disk & 0.0 & 1.0 & 0.33 & 0.33 & 0.0 \\
deep2\textunderscore8523 & Clumpy Disk & 0.5 & 1.0 & 0.17 & 0.83 & 0.0 \\
deep2\textunderscore7186 & Clumpy Disk + Spheroid & 0.6 & 0.6 & 0.4 & 0.8 & 0.0 \\
ers2\textunderscore10675 & Spheroid & 0.67 & 0.33 & 0.0 & 0.0 & 0.0 \\
ers2\textunderscore13558 & Irregular & 0.33 & 0.0 & 0.67 & 0.0 & 0.0 \\
deep4\textunderscore8133 & Merger & 0.4 & 0.6 & 0.6 & 0.8 & 0.6 \\
\enddata
\end{deluxetable}

\begin{deluxetable}{ccccccccccc}[ht]
\tablecaption{Raw catalog votes from \citetalias{K15} for the Merger deep4\textunderscore8133, whose vote fractions are shown in the sixth row of Table \ref{ClassTable}.  Each row represents the votes of a specific classifier.  For the Interaction Class, 0 = No interaction, 1 = Merger, 2 = Interaction within the segmentation map, 3 = Interaction outside the segmentation map, and 4 = Non-interacting companion.  In other columns, a 1 is shown if the classifier voted for the given class while a 0 is given otherwise.  The clumpiness flags (e.g. C1P0) shown here are used to derive the vote fraction $f_{Clumpy}$.}
\tablecolumns{7}
\tablenum{3}
\label{ClassTableRaw}
\tablewidth{0pt}
\tablehead{
\colhead{Classifier ID} &
\colhead{Spheroid} &
\colhead{Disk} & \colhead{Irregular} & \colhead{Interaction Class} & \colhead{C1P0} & \colhead{C1P1} & \colhead{C1P2} & \colhead{C2P0} & \colhead{C2P1} & \colhead{C2P2}
}
\startdata
5 & 1 & 1 & 1 & 1 & 0 & 1 & 0 & 0 & 0 & 0 \\
28 & 0 & 0 & 1 & 1 & 1 & 0 & 0 & 0 & 0 & 0 \\
35 & 1 & 1 & 0 & 2 & 0 & 1 & 0 & 0 & 0 & 0 \\
39 & 0 & 0 & 1 & 1 & 0 & 1 & 0 & 0 & 0 & 0 \\
43 & 0 & 1 & 0 & 3 & 0 & 0 & 0 & 0 & 0 & 0 \\
\enddata
\end{deluxetable}

\begin{figure}[ht]
\centering
\includegraphics[scale=0.5]{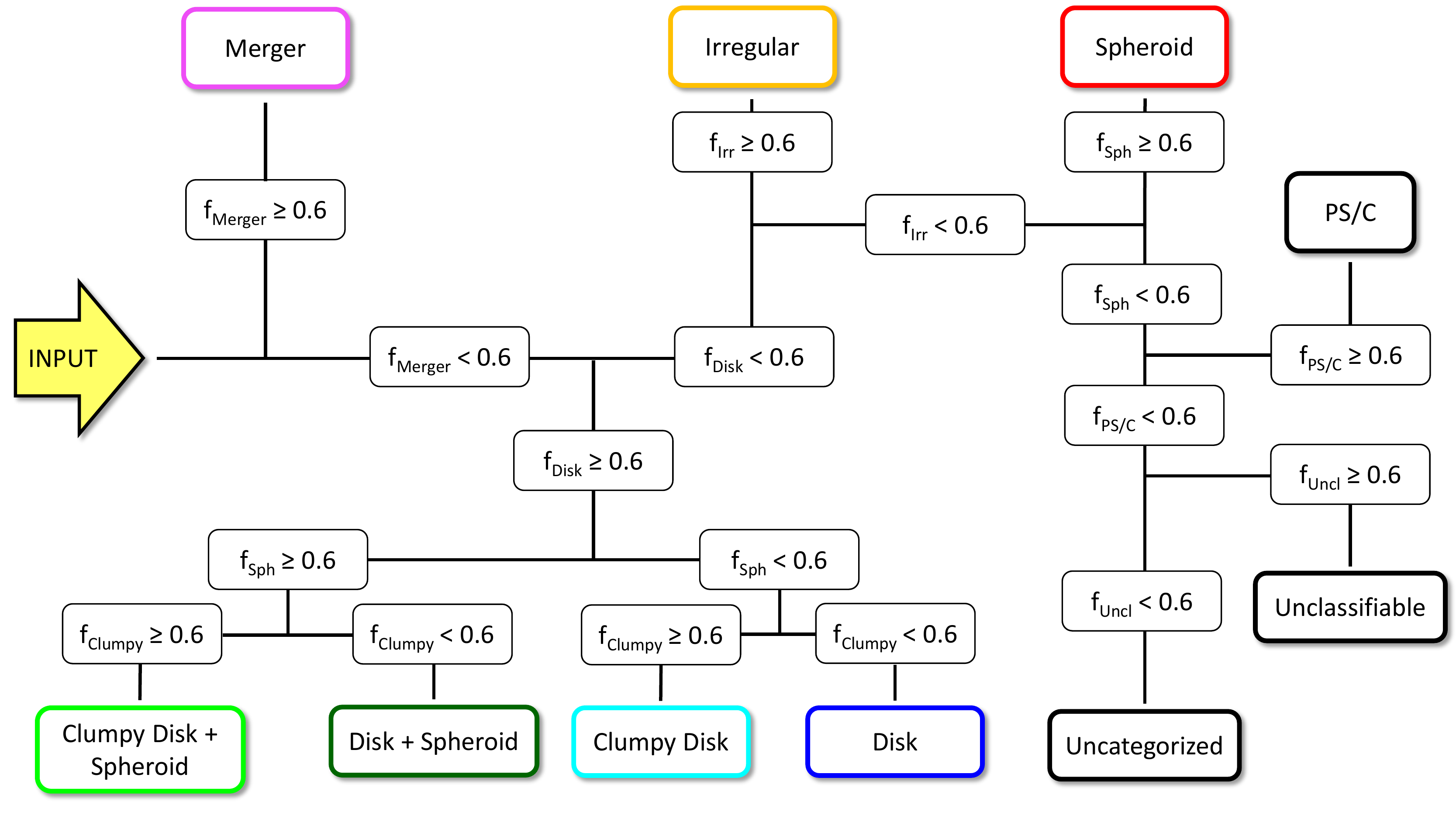}
\caption{A decision tree elucidating the conversion from the \citetalias{K15} vote fractions ($f$) to the ten mutually exclusive classes used in this work.  All vote fractions are taken from the fractional catalog with the exception of $f_{Clumpy}$, which we calculate using the raw catalog.}
\label{DecisionTree}
\end{figure}

In this section we describe our conversion from the \citetalias{K15} catalogs to the discrete and simplified scheme used in this work.  For each object, the \citetalias{K15} classifiers were instructed to choose at least one Main Morphology Class (which includes Disk, Spheroid, and Irregular/Peculiar) and one option from the Interaction Class.  The Interaction Class has five options:  no interaction, non-interacting companion, interaction outside/within the segmentation map, and merger.  Classifiers were encouraged to choose multiple main morphology classes if applicable, and were allowed to select any number of the structural, quality, k-correction, or clumpiness/patchiness flags.  

The raw catalog contains the votes made by each individual classifier for each of the objects that they were assigned.  The fractional catalog contains one entry for each object, and for each class a vote fraction is given which represents the fraction of classifiers who voted for that class.  Each object therefore has a set of `vote fractions' representing the fraction of classifiers who voted for each category. 

We show the vote fractions for several example galaxies of different classes in Table \ref{ClassTable}, while in Table \ref{ClassTableRaw} we show the raw catalog votes for one of the galaxies from Table \ref{ClassTable}.  All of the vote fractions we use for classification are taken from the fractional catalog, with the exception of $f_{Clumpy}$.  In the case of $f_{Clumpy}$, we form our own vote fraction because the vote fractions for the clumpiness flags given in the fractional catalog do not account for single classifiers who checked multiple flags.  To account for this, we use the raw catalog to calculate $f_{Clumpy}$ instead as the number of classifiers who voted for a non-zero number of clumps divided by the total number of classifiers.  

Our full class assignment process is shown in Figure \ref{DecisionTree}.  We consider a vote fraction (denoted by $f$) threshold of 0.6 (corresponding to $60\%$ of classifiers voting for a category) to represent a majority vote, and use this threshold to assign galaxies to different classes.  We find that higher thresholds exclude most of the galaxies that would be classified as Mergers in our nominal scheme, preventing a meaningful analysis of their properties.  Adopting a more inclusive threshold of 0.5 has virtually no impact on our results.  

\section{Impact of Size and Magnitude Biases on Morphology}
\label{Section:BiasCheck}

\begin{figure*}[ht]
    \centering
    \includegraphics[scale=0.4]{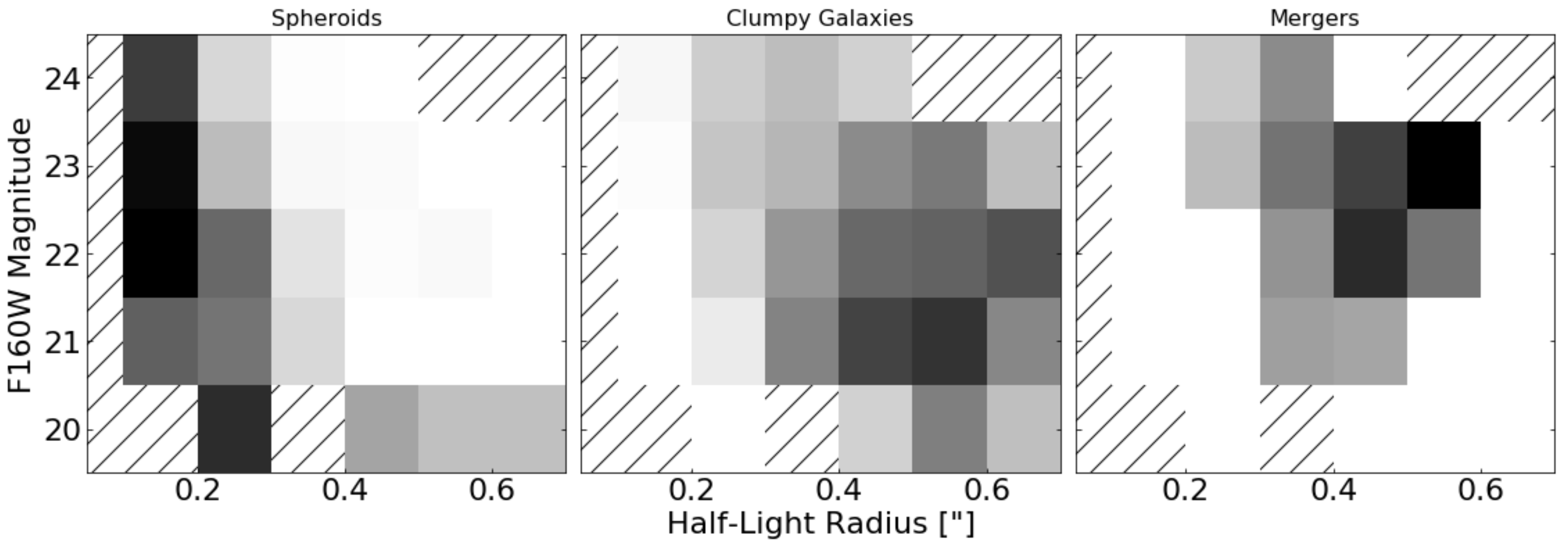}
    \caption{Number fraction for different classes among our entire GOODS-S sample ($z \sim 1$ and $z \sim 2$) across different bins in \textit{H}-band (F160W) magnitude and half-light radius (in arcseconds).  We use these plots to examine potential biases in the detection of Spheroids (left), Clumpy Disks and Clumpy Disk + Spheroids (Clumpy galaxies collectively, middle) and Mergers (right).  Size and magnitude measurements are taken from \citet{vanderWel2012CANDELSStructuralParams}.  Darker shades indicate higher relative fractions.  The shading in each individual plot is normalized and so does not represent the absolute fractions.  Bins with $<5$ objects are unshaded and marked by diagonal lines.  Spheroids prefer smaller galaxies whereas Clumpy galaxies and possibly Mergers may prefer larger galaxies.  We expect Spheroids to be compact and disks to be extended, so this is unlikely to be a classification bias.  No morphology shows a clear bias in magnitude.  Overall, our classifications appear to be largely unaffected by biases arising from size or apparent brightness.}
    \label{BiasCheck}
\end{figure*}

In this section, we investigate potential biases in our sample that stem from variations in size, depth, and apparent magnitude, and discuss the possible impact of cosmological surface brightness dimming on our results.  We show in Figure \ref{BiasCheck} the variation in number fraction of Spheroids, Clumpy galaxies (Clumpy Disks and Clumpy Disk + Spheroids), and Mergers in both size (half-light radius, in arcseconds) and \textit{H}-band apparent magnitude for our entire GOODS-S sample ($z \sim 1$ and $z \sim 2$).  The size and magnitude measurements are taken from \citet{vanderWel2012CANDELSStructuralParams}.  We do not separate by redshift because the physical resolution scale varies little ($\approx 5\%$) across the redshift range $0.8 < z < 2.5$.  

In Figure \ref{BiasCheck}, Spheroids show a strong preference for smaller sizes while showing little to no preference in magnitude.  This is reassuring, as one might expect fainter galaxies to be preferentially classified as Spheroids.  Since we expect Spheroids to be generally more compact \citep[see][and references therein]{vanderWel2014GalaxyMassSizeEvol}, their preference for smaller sizes is unsurprising and does not necessarily constitute a bias.  Clumpy galaxies appear to prefer larger sizes and show no magnitude preference.  We expect disk galaxies to be generally larger in size, however \citep[see again][]{vanderWel2014GalaxyMassSizeEvol}, and it may simply be easier to `fit' a clump into a larger disk.  Mergers may have some preference for larger sizes, but have no clear magnitude preference.  Overall, we do not find clear evidence of significant biases arising from size or apparent magnitude.  The \textit{H}-band magnitude cut used by \citetalias{K15} means that our sample consists of relatively bright objects; this likely helps to alleviate many of the expected visual biases arising due to size or apparent magnitude.  

Another potential bias arises due to cosmological surface brightness dimming, which has a strong redshift dependence ($\propto (1+z)^{-4}$).  This may render disk substructures (e.g., clumps) and characteristic merger features (e.g., tidal tails) less prominent at higher redshifts.  To test this, we repeated our analysis using only galaxies in the `deep' region of GOODS-S for our $z \sim 1$ and $z \sim 2$ samples.  We find that, although the fraction of both Clumpy Disks and Clumpy Disk + Spheroids increases (indicating that clump detection is indeed affected by depth), our main conclusions are largely unchanged.  Our Merger fractions are also unaffected.  The main limitation of this method is that the deep region of GOODS-S is only $\sim 0.5$ magnitudes deeper on average, while the dimming at $z = 2$ reduces the surface brightness by $\sim 5$ magnitudes.  While not conclusive, this suggests that cosmological dimming may not have a strong impact on our results.  

\bibliography{Bibliography}
\bibliographystyle{aasjournal}

\end{document}